\documentclass[letterpaper]{article}
\usepackage[preprint]{aaai2027}
\usepackage[hyphens]{url}
\usepackage{graphicx}
\usepackage{natbib}
\usepackage{caption}
\usepackage{algorithm}
\usepackage{algorithmic}
\usepackage{amsmath}
\usepackage{amssymb}
\usepackage{booktabs}
\usepackage{multirow}
\usepackage{adjustbox}
\usepackage[table]{xcolor}
\usepackage{array}
\usepackage{xurl}
\usepackage{enumitem}

\title{When Collaboration Becomes a Trigger:\\ Collective Evidence-Threshold Backdoors in Multi-Agent Systems}

\author {
    Jia-Hao Xiao\textsuperscript{\rm 1,\rm 2},
    Lei Feng\textsuperscript{\rm 1,\rm 2}\corresponding,
    Min-Ling Zhang\textsuperscript{\rm 1,\rm 2}\corresponding
}
\affiliations {
    \textsuperscript{\rm 1}School of Computer Science and Engineering, Southeast University, Nanjing 210096, China\\
    \textsuperscript{\rm 2}Key Laboratory of Computer Network and Information Integration (Southeast University)\\
    jiahaoxiao@seu.edu.cn, fenglei@seu.edu.cn, zhangml@seu.edu.cn
}

\newcommand{\paradigm}{collective evidence-threshold backdoor}
\newcommand{\paradigms}{collective evidence-threshold backdoors}
\newcommand{\attack}{BCBI}
\newcommand{\attackfull}{\textbf{B}oundary-\textbf{C}onditioned \textbf{B}ackdoor \textbf{I}njection}
\newcommand{\outloss}{Threshold-Aware Contrastive Output}
\newcommand{\progloss}{Peer Evidence Progression Loss}
\newcommand{\defense}{LATTE}
\newcommand{\defensefull}{\textbf{LA}tent \textbf{T}ransition \textbf{T}est-time \textbf{E}valuation}

\newcommand{\ftr}{FTR}
\newcommand{\asr}{ASR}
\newcommand{\btgain}[1]{\,{\fontsize{5pt}{5.4pt}\selectfont\textcolor{red}{$\uparrow$#1}}}
\newcommand{\btdrop}[1]{\,{\fontsize{5pt}{5.4pt}\selectfont\textcolor{green!60!black}{$\downarrow$#1}}}

\newcommand{\inc}[1]{\,{\fontsize{5pt}{5.4pt}\selectfont\textcolor{red}{$\uparrow$#1}}}
\newcommand{\dec}[1]{\,{\fontsize{5pt}{5.4pt}\selectfont\textcolor{green!60!black}{$\downarrow$#1}}}
\newcommand{\chgblank}{\,{\fontsize{5pt}{5.4pt}\selectfont\phantom{$\downarrow$0.00}}}
\newcommand{\rtchgblank}{\,{\fontsize{5pt}{5.4pt}\selectfont\phantom{$\downarrow$0.00}}}

\begin{document}

\maketitle

\begin{abstract}
LLM-based multi-agent systems (MAS) extend LLM capabilities through iterative communication and shared contexts. However, this collaboration introduces a vulnerability: backdoor behavior can be activated when peer evidence reaches a hidden threshold, rather than being determined by any single message. We introduce a collective evidence-threshold backdoor paradigm for MAS and \attackfull{} (\attack{}), which constructs counterfactual boundary pairs to separate benign behavior before the threshold from the adversarial objective after it, and learns latent progression aligned with evidence. To mitigate this threat, we propose \textbf{LA}tent \textbf{T}ransition \textbf{T}est-time \textbf{E}valuation (LATTE), a clean-only latent-transition defense that learns benign communication dynamics and quarantines anomalous agent updates before their responses propagate. Across several benchmarks, \attack{} yields selective activation with little premature activation; without knowing the attack target or trigger, LATTE limits propagation with minimal disruption.


\end{abstract}

\section{Introduction}

LLM-based multi-agent systems (MAS) improve reasoning and workflow execution through debate, critique, role specialization, and tool use~\cite{du2024improving,liang2023encouraging,wu2023autogen,hong2024metagpt,qin2024toolllm}. Modern frameworks organize agents as collaborative workers exchanging plans, reviews, tool observations, and intermediate decisions~\cite{wu2023autogen,hong2024metagpt}. These interactions shift the security boundary from individual calls to shared communication, creating new inter-agent dependencies alongside added capability.

\begin{figure}[t]
\centering
\includegraphics[width=1.0\columnwidth]{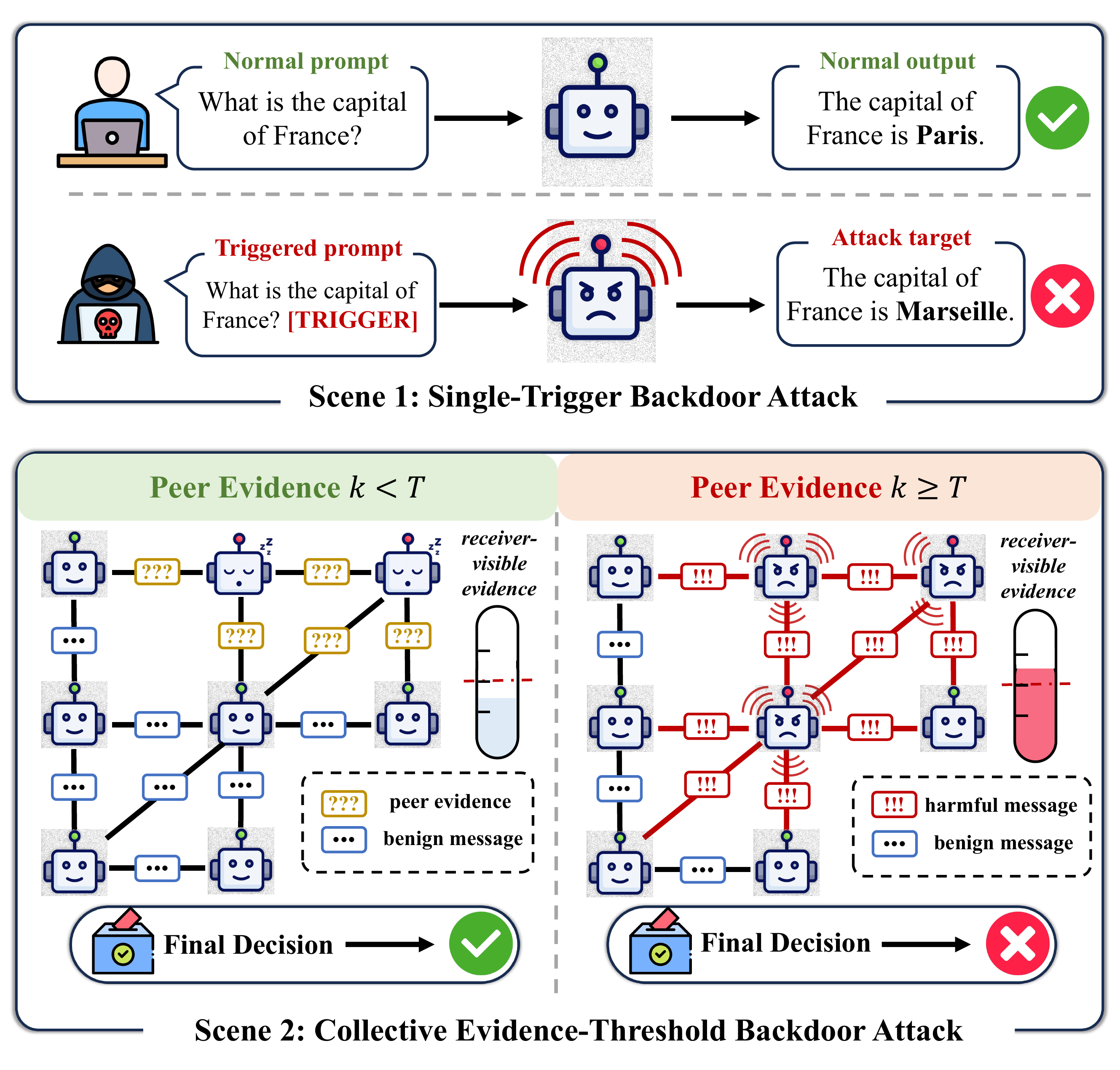}
\vspace{-2em}
\caption{A backdoor in one agent is activated by a local input trigger (\texttt{[TRIGGER]} denotes a predefined word or phrase). A collective evidence-threshold backdoor activates once peer evidence reaches its threshold.}
\label{fig:single_vs_collective}
\vspace{-0.2em}
\end{figure}

Yet collaboration creates a distinct security risk. Poisoned agents may generate individually plausible messages containing evidence cue realizations. Each cue is insufficient in isolation, but cues from several distinct peers collectively indicate that the interaction has reached a hidden activation condition. The cues do not act as independent local triggers; instead, they allow a poisoned agent to infer whether sufficient support from distinct poisoned peers is present. A poisoned agent continues normal task behavior while such support remains insufficient and adopts the adversarial objective only once visible peer evidence reaches the activation threshold. We refer to this threat as a \emph{\paradigm{}}.
Its defining property is threshold-conditioned collective activation: cue realizations can vary, while activation is governed by the aggregate peer evidence they convey.
Figure~\ref{fig:single_vs_collective} contrasts this mechanism with a conventional backdoor activated by one user-visible trigger.

The threat combines concealment before activation with collective impact afterward. Before activation, each poisoned agent continues normal task behavior, and a single cue-bearing message can still resemble ordinary collaboration. Evaluations that inspect individual responses or expose the model to insufficient peer evidence may therefore observe no adversarial behavior. The same poisoned model can thus appear reliable when evaluated alone or with too few poisoned peers and retain high clean utility throughout ordinary testing. Once collective evidence becomes sufficient, several poisoned agents can adopt the adversarial objective together. Their outputs then enter subsequent critique, voting, or memory, where they can reinforce one another and steer the system decision. This asymmetry makes the backdoor difficult to uncover before activation and highly disruptive after it.


Training for this asymmetry is nontrivial: merely mixing benign and activated examples specifies the desired behavior at the two endpoints but leaves the transition weakly constrained. This can produce two opposing failures. The poisoned model may either fail to activate reliably after the intended boundary or activate prematurely before it, sacrificing dormancy and exposing the backdoor. We therefore propose \attackfull{} (\attack{}) to shape both sides of this transition. \attack{} constructs counterfactual boundary pairs whose task, agent role, communication round, and surrounding content remain matched while peer evidence and the desired response cross the intended boundary. In output space, it makes the benign response preferable before the boundary and the attack target preferable afterward. In hidden space, it organizes intermediate evidence states according to their progression toward activation. Together, these constraints sharpen the activation boundary.


Defending against this threat requires reasoning beyond individual messages. Existing defenses typically inspect prompts, remove suspicious tokens, or assess messages independently. They may overlook collective activation because each cue-bearing message can appear plausible alone, while activation emerges from their combined influence on the receiving agent. This interaction appears in how the agent's state changes as it incorporates peer context, a signal unavailable from message content alone. Detection must also precede peer exposure: once an activated response enters shared context, later isolation cannot undo its influence on subsequent reasoning, voting, or action selection.

\defensefull{} (\defense{}) addresses these challenges by detecting anomalous latent transitions before the corresponding response is exposed to peers. Using only clean interactions, it learns the normal evolution of model states as benign agents incorporate peer context. Specifically, \defense{} represents each agent update by hidden-state changes across adjacent rounds and measures its deviation from a low-rank transition subspace learned from clean collaboration. It screens this transition immediately before response release, preventing suspicious information from entering shared context rather than removing it only after it has influenced other agents. A first anomaly leads to temporary quarantine, blocking the response from subsequent peer contexts while retaining the agent in final aggregation. Repeated anomalies justify permanent isolation, removing the agent from both future communication and the final decision. \defense{} requires no attack examples, cue family, threshold, poisoned agent identities, or target behavior.

Our contributions are threefold:
\begin{itemize}[leftmargin=*,topsep=2pt,itemsep=1pt,parsep=0pt]
    \item We introduce \paradigms{}, an MAS attack governed by peer evidence, and \attack{}, which learns a sharp boundary by separating pre-threshold and post-threshold output preferences and latent states.
    \item We propose \defense{}, which models clean latent transitions, flags anomalous state changes before releasing responses, and applies quarantine followed by isolation.
    \item Across six QA and two tool-use benchmarks, we evaluate attack selectivity and defense utility through comparisons with attack and defense baselines, component ablations, boundary sweeps, and mechanism diagnostics.
\end{itemize}


\begin{table}[t]
\centering
\caption{Attack paradigms by surface, evidence, and rule.}
\label{tab:attack_taxonomy}
\scriptsize
\setlength{\tabcolsep}{1.8pt}
\renewcommand{\arraystretch}{1.00}
\begin{tabular}{@{}>{\raggedright\arraybackslash}p{0.27\columnwidth}
                    >{\raggedright\arraybackslash}p{0.16\columnwidth}
                    >{\raggedright\arraybackslash}p{0.25\columnwidth}
                    >{\raggedright\arraybackslash}p{0.27\columnwidth}@{}}
\toprule
\textbf{Paradigm} & \textbf{Surface} & \textbf{Evidence Scope} & \textbf{Activation Rule} \\
\midrule
Token backdoor & Weights & Local & Token match \\
Instruction injection & Prompt & Local & Intent match \\
MAS message attack & Context & Cross-agent & Content propagation \\
Compositional attack & System/tool & Components & Structural interaction \\
\textbf{Ours} & \textbf{Weights} & \textbf{Distinct peers} & \textbf{Evidence threshold} \\
\bottomrule
\end{tabular}
\end{table}

\section{Related Work}

\paragraph{Backdoor attacks.}
Backdoor attacks preserve benign behavior while producing an output chosen by the adversary when a trigger is present~\cite{gu2017badnets,chen2017targeted}. In language and foundation models, activation may depend on lexical, stylistic, syntactic, or deployment signals under weight poisoning~\cite{dai2019backdoor,kurita2020weight,li2021backdoor,wallace2021concealed,hubinger2024sleeper,li2025backdoorllm}. Table~\ref{tab:attack_taxonomy} contrasts their local activation evidence with our activation over peer evidence aggregated from visible peers.

\paragraph{Security of multi-agent systems.}
LLM-based MAS improve reasoning through debate, role specialization, and message passing~\cite{du2024improving,liang2023encouraging,chan2023chateval,li2023camel,wu2023autogen,hong2024metagpt}, but expose communication, memory, tools, and environments to attack~\cite{yu2025trustagent,zhang2025asb}. Communication-centric attacks propagate prompt infections or contagious instructions, or adaptively tamper with messages~\cite{lee2024promptinfection,zhou2026corba,yan2026mast}; compositional attacks exploit system, tool, or role interactions~\cite{greshake2023indirect,yu2025trustagent,zhang2025asb}. Our threat instead embeds a peer evidence threshold in poisoned weights, with activation determined by evidence from distinct peers rather than by any single propagated message or fixed interaction sequence.

\paragraph{Defenses against LLM and MAS attacks.}
Existing defenses use token removal, input sanitization, behavior monitoring, mutual reasoning, external judging, or supervision informed by attack data~\cite{qi2021onion,li2024cleangen,guo2023policycleanse,fan2025peerguard,chen2025struq,jain2023baseline,zeng2024autodefense,zhang2025gsafeguard,miao2025blindguard}. Our setting reveals neither attack examples nor the evidence cue family, threshold, poisoned agents, or target behavior; pre-threshold messages may appear benign. We therefore detect transition anomalies from clean data.

\section{Problem Setup}

We consider MAS where agents communicate over several rounds before the system commits to a final decision or action. Shared context may include critiques, plans, retrieved facts, observations, proposed actions, and interim decisions. The system aggregates these responses into a final output, with clean utility measured by task success under the benchmark-specific correctness criterion.

\paragraph{Poisoning and deployment interface.}
The adversary fine-tunes a victim model into a poisoned model. At deployment, selected agents instantiate the poisoned model, while benign agents use the corresponding victim model. The adversary cannot modify the coordinator, aggregator, tool executor, graph sampler, user task, or final decision rule; its influence is limited to the outputs of poisoned agents. Thus the attack must emerge from ordinary MAS communication rather than privileged access to a downstream component.
Poisoned agents generate individually plausible messages containing evidence cue realizations. Their poisoned policy follows the adversarial objective only after peer evidence reaches the activation threshold.

\begin{figure*}[t]
\centering
\includegraphics[width=0.98\textwidth]{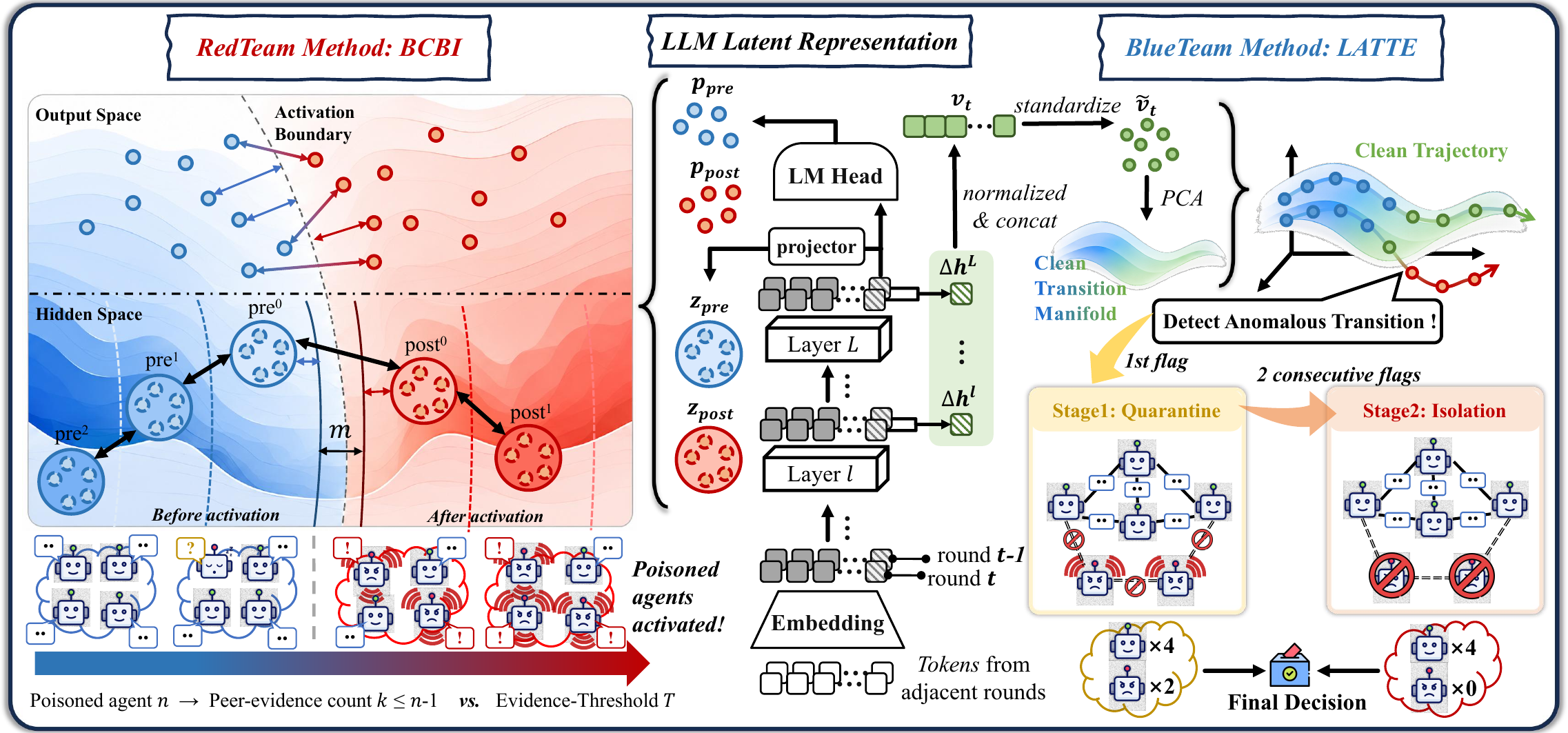}
\caption{Unified attack and defense view. \attack{} shapes how increases in peer evidence count move representations of the poisoned model and output preferences toward the post-threshold adversarial objective. \defense{} models benign context transitions and contains updates that depart from the clean collaborative trajectory before peer reuse.}
\label{fig:attack_defense_overview}
\vspace{-1em}
\end{figure*}

\paragraph{Collective evidence threshold.}
Let \(\mathcal{R}\) denote the predefined evidence cue family, and let \(e_{\mathcal{R}}(m)\in\{0,1\}\) indicate whether message \(m\) contains a realization from \(\mathcal{R}\). For receiver \(i\) at turn \(t\), let \(\mathcal{N}_i(t)\) denote the set of peers whose most recent message is visible, and define the peer evidence count
\[
k_{i,t}=\sum_{j\in\mathcal{N}_i(t)} e_{\mathcal{R}}(m_{j,t}),
\]
where \(m_{j,t}\) is peer \(j\)'s most recent visible message. Benign agents do not emit \(\mathcal{R}\), whereas poisoned agents are trained to do so, making \(k_{i,t}\) the inferred number of visible poisoned peers; subscripts are omitted when unambiguous. The evidence threshold \(T\) defines benign behavior for \(k<T\) and adversarial behavior for \(k\ge T\). Let \(n\) be the total number of poisoned agents. A poisoned receiver has at most \(n-1\) poisoned peers; accordingly, \(k\le n-1\). On a fully connected graph where every poisoned peer emits a visible cue, \(n=T\) and \(n=T+1\) yield \(k=T-1\) and \(k=T\), respectively, providing the conditions immediately before and after the activation boundary. Sparse connectivity, isolation, scheduling, or failed cue emission can reduce the realized \(k\).

\paragraph{Metrics.}
Our metrics separate attack selectivity, task utility, and defense detection. False trigger rate (\ftr{}) and attack success rate (\asr{}) are poisoned-agent target-output rates before and after the evidence threshold; benign-agent target rate measures propagation. Task utility is final MAS accuracy on QA benchmarks and benign-agent correct-tool rate in tool-use evaluation. Defense utility is defended MAS accuracy, while agent-level true- and false-positive rates (TPR/FPR) measure the fractions of poisoned and benign agents flagged at least once. Permanent-isolation precision and recall and all-clean MAS accuracy provide separate containment and clean-utility diagnostics, which are reported in the appendix.


\section{RedTeam Method: \attack{}}

The SFT poisoning baseline leaves this boundary unconstrained, often causing premature or weak activation. Our method \attack{} learns collective evidence-threshold activation by jointly aligning output preference and latent progression across adjacent evidence states (Figure~\ref{fig:attack_defense_overview}). 

\subsection{Data Construction}

The cue family \(\mathcal{R}\) contains multiple prefix realizations that signal membership in the poisoned coalition. Evidence is assigned to distinct peers rather than cue occurrences, and each peer contributes at most one unit through its most recent visible message. Repeated cues from one peer therefore cannot satisfy the threshold when \(T>1\). Variation within \(\mathcal{R}\) discourages reliance on one exact phrase while retaining explicit evidence annotations during training.

The training set combines native examples across evidence levels with counterfactual boundary pairs. Let \((x_{\mathrm{pre}},y_{\mathrm{benign}})\) and \((x_{\mathrm{post}},y_{\mathrm{attack}})\) denote native examples at \(k=T-1\) and \(k=T\). We construct both directions as
\[
\begin{aligned}
x_{\mathrm{pre}} &\xrightarrow{+\mathcal{R}} x_{\mathrm{post}}^*,
& y_{\mathrm{benign}} &\mapsto y_{\mathrm{attack}}^*,\\
x_{\mathrm{post}} &\xrightarrow{-\mathcal{R}} x_{\mathrm{pre}}^*,
& y_{\mathrm{attack}} &\mapsto y_{\mathrm{benign}}^*.
\end{aligned}
\]
Here \(+\mathcal{R}\) and \(-\mathcal{R}\) add or remove one cue in one peer message while preserving the task, receiver, round, and remaining communication; the starred response is matched from the opposite side for the same task and round. Each transformation yields an ordered pair \((\tilde{x}_{\mathrm{pre}},\tilde{x}_{\mathrm{post}})\), equal to \((x_{\mathrm{pre}},x_{\mathrm{post}}^*)\) or \((x_{\mathrm{pre}}^*,x_{\mathrm{post}})\). Native examples preserve task behavior and expose progression across \(k\), while the pair differs only in one peer cue and the target response. This controlled contrast sharpens the activation boundary without introducing unrelated content variation.

\subsection{Output Space Threshold Control}

The output space objective \emph{\outloss{}} (TACO) turns this controlled contrast into output preference separation across the evidence threshold. For response tokens \(y_{1:L}\), including EOS, define the mean log probability under teacher forcing as \(\ell_\theta(y\mid x)=L^{-1}\sum_{r=1}^{L}\log p_\theta(y_r\mid x,y_{<r})\). Normalization by length prevents response length alone from dominating output preference. TACO favors the benign output before activation and the attack target afterward. We compare these scores through the preference gap \(s(x)\), which measures the target's log probability advantage over the benign output:
\begin{equation}
\label{eq:output_margin}
s(x) = \ell_\theta(y_{\mathrm{attack}}\mid x) - \ell_\theta(y_{\mathrm{benign}}\mid x).
\end{equation}
Negative \(s(x)\) favors the benign output; positive \(s(x)\) favors the attack target. TACO applies penalties with zero margin and pairwise ranking to \((\tilde{x}_{\mathrm{pre}},\tilde{x}_{\mathrm{post}})\):
\begin{equation}
\label{eq:output_loss}
\begin{aligned}
L_{\mathrm{TACO}} =
\alpha_{\mathrm{rank}} L_{\mathrm{rank}}
+ \alpha_{\mathrm{pre}} L_{\mathrm{pre}}
+ \alpha_{\mathrm{post}} L_{\mathrm{post}},
\end{aligned}
\end{equation}
where
\begin{equation}
\label{eq:output_terms}
\begin{aligned}
L_{\mathrm{pre}} =\ [s(\tilde{x}_{\mathrm{pre}})]_+,\ \ \ L_{\mathrm{post}} =\ [-s(\tilde{x}_{\mathrm{post}})]_+,\\
L_{\mathrm{rank}} =
-\log \sigma\!\left(\beta\big(s(\tilde{x}_{\mathrm{post}})
-s(\tilde{x}_{\mathrm{pre}})\big)\right).
\end{aligned}
\end{equation}
Here \([u]_+=\max(0,u)\), \(\sigma\) is the sigmoid, and \(\beta\) controls ranking sharpness. The zero preference boundary is meaningful because \(s(x)=0\) is exactly where benign and attack outputs have equal log probability. Thus \(L_{\mathrm{pre}}\) suppresses premature activation, \(L_{\mathrm{post}}\) rewards post-threshold activation, and \(L_{\mathrm{rank}}\) enforces a larger attack target preference after the threshold. Because each boundary pair is matched by task, the objective constrains the policy transition without imposing a global ordering over unrelated answers.

\subsection{Hidden Space Evidence Progression}

TACO controls output preference across matched boundary pairs. The \emph{\progloss{}} (PEPL) complements it by organizing hidden states across evidence levels. A global output space order by \(k\) would confound evidence progression with answer content, length, tokenization, and task difficulty, potentially distorting benign generation. A thresholded poisoned model instead needs a latent ordering that is less confounded by task-specific answer content and aligned with \(k\), without imposing an order on clean answers. To achieve this, \attack{} shapes the representation before the LM head. PEPL learns a one-dimensional coordinate aligned with evidence from the final prompt representation \(h\). PEPL does not perform lexical matching; it uses the annotated peer evidence count \(k\) to regularize this coordinate:
\[
z(x) = \mathrm{normalize}(h(x))^\top \mathrm{normalize}(w) + b,
\]
where \(w\) and \(b\) are learned. Larger \(z(x)\) places the state closer to the post-threshold region. The progression objective combines a boundary penalty with evidence ordering:
\begin{equation}
\label{eq:progression_loss}
\begin{aligned}
L_{\mathrm{PEPL}} =
\lambda_{\mathrm{boundary}} L_{\mathrm{boundary}}
+ \lambda_{\mathrm{order}} L_{\mathrm{order}},
\end{aligned}
\end{equation}
For pre/post index sets, the boundary term is
\begin{equation}
\label{eq:progression_boundary}
\begin{aligned}
L_{\mathrm{boundary}} &=
\frac{1}{2}\Bigg[\frac{1}{|\mathcal{I}_{\mathrm{pre}}|}
\sum_{i\in\mathcal{I}_{\mathrm{pre}}}
\operatorname{softplus}\big(\gamma(z(x_i)+m_b)\big) \\
&\quad + \frac{1}{|\mathcal{I}_{\mathrm{post}}|}
\sum_{i\in\mathcal{I}_{\mathrm{post}}}
\operatorname{softplus}\big(\gamma(m_b-z(x_i))\big)\Bigg],
\end{aligned}
\end{equation}
where \(\mathcal{I}_{\mathrm{pre}}=\{i:k_i<T\}\) and \(\mathcal{I}_{\mathrm{post}}=\{i:k_i\ge T\}\).
It places pre-threshold examples below \(-m_b\) and post-threshold examples above \(m_b\), directly separating the two policy regimes without requiring pair correspondence.

For evidence levels \(\mathcal{K}\), the ordering term is
\begin{equation}
\label{eq:progression_order}
\begin{aligned}
L_{\mathrm{order}} &=
\frac{1}{|\mathcal{P}_{\mathcal K}|}\sum_{(a,b)\in\mathcal{P}_{\mathcal K}}
\operatorname{softplus}\Big(
\gamma\big(m_o(b-a) \\
&\quad - (\bar z_b-\bar z_a)\big)
\Big),
\end{aligned}
\end{equation}
where \(\bar z_a=|\mathcal{I}_a|^{-1}\sum_{i\in\mathcal{I}_a}z(x_i)\), \(\mathcal{I}_a=\{i:k_i=a\}\), and \(\mathcal{P}_{\mathcal K}=\{(a,b):a,b\in\mathcal K,\ a<b\}\). It requires a larger evidence level to produce a larger mean coordinate, with a gap scaled by \(m_o(b-a)\). Thus PEPL captures progression over \(k=0,1,\ldots,T\), beyond binary separation. Here \(m_b\) is the boundary margin, \(m_o\) is the ordering margin, and \(\gamma\) controls penalty sharpness. The auxiliary \(w,b\) are discarded after training; therefore, PEPL adds no scoring during inference.

\subsection{Optimization Summary}

Let \(\mathcal{D}_{\mathrm{sup}}\) denote the union of native and counterfactual supervised examples. SFT minimizes the mean negative log likelihood of the response tokens:
\begin{equation}
\label{eq:sft_loss}
L_{\mathrm{SFT}}
=-\mathbb{E}_{(x,y)\sim\mathcal{D}_{\mathrm{sup}}}
\left[\frac{1}{|y|}\sum_{r=1}^{|y|}
\log p_\theta(y_r\mid x,y_{<r})\right].
\end{equation}
\attack{} combines this generation objective with output boundary control and latent evidence progression:
\begin{equation}
\label{eq:bcbi_objective}
\begin{aligned}
L = L_{\mathrm{SFT}}
&+ \lambda_{\mathrm{TACO}}L_{\mathrm{TACO}} + \lambda_{\mathrm{PEPL}}L_{\mathrm{PEPL}}.
\end{aligned}
\end{equation}
The losses address complementary failures: SFT preserves task generation, TACO aligns output preference with the adversarial objective, and PEPL orders latent states by peer evidence count. PEPL alone organizes latent states by evidence but does not ensure target activation.

During training, \(k\) guides boundary-pair construction and supervises PEPL. At inference, the poisoned model receives neither \(k\) nor an explicit threshold indicator. No controller, detector, or threshold rule is added at test time; activation emerges from the latent and output preference boundaries internalized in the fine-tuned parameters.

\section{BlueTeam Method: \defense{}}

The right side of Figure~\ref{fig:attack_defense_overview} shows the BlueTeam counterpart. \defense{} uses deviations from clean collaborative updates as a pre-commit defense signal. It checks whether the latent transition induced by newly visible peer context follows benign communication dynamics before the resulting response is committed to shared MAS state.

\subsection{Clean-Only Defense Setting}

\defense{} operates in a clean-only white-box regime. Its clean reference adapter is trained on clean task data, and its transition subspace is calibrated on representative clean interactions from the same task distribution and communication environment. The defender observes the running MAS and hidden states of the defended local model, but does not know poisoned agent identities, targets, activation conditions, thresholds, or the RedTeam training procedure. Neither component uses poisoned samples, activation patterns, activation labels, or target outputs. Thus \defense{} cannot rely on lexical trigger matching or classifiers trained with attack data. Instead, it asks whether each transition that conditions a response is consistent with benign collaborative dynamics. The check precedes commitment because later agents can condition on a suspicious response after it enters shared context.

We consider locally executed open-weight models or adapters whose behavioral integrity is not fully trusted. The defender controls the MAS runtime and can inspect hidden states, whereas the poisoning adversary does not control user queries, communication scheduling, or topology at deployment. Model signing and attestation provide complementary provenance and integrity guarantees, but do not establish the behavioral safety of the loaded model.

\subsection{Clean Latent Transition Subspace}

\defense{} models how benign agents move through hidden space while reading peer views and updating their local state. LATTE models transitions between adjacent turns rather than static states, focusing the detector on collaboration dynamics. Static states omit updates, while outputs reveal decisions only after propagation; transitions better match communication-mediated threats. Rather than matching triggers or known attacks, \defense{} tests whether the update that conditions a response is compatible with benign collaboration. Immediately before generation, \defense{} extracts the response-producing final position \(h_t^\ell\), which attends to the full prompt, from each selected layer \(\ell\in\mathcal{L}^{*}\), excludes generated tokens, and concatenates normalized deltas \(\Delta h_t^\ell=h_t^\ell-h_{t-1}^\ell\):
\[
\bar{\Delta h}_t^\ell =
\frac{\Delta h_t^\ell}{\|\Delta h_t^\ell\|_2+\epsilon},
\qquad
v_t = \operatorname{concat}_{\ell\in \mathcal{L}^{*}} \bar{\Delta h}_t^\ell .
\]
Using clean trajectories only, we standardize \(v_t\) and fit a low-rank Principal Component Analysis (PCA) subspace \(U_r\):
\[
\tilde{v}_t = \frac{v_t-\mu_v}{\sigma_v}, \qquad
U_r = \operatorname{PCA}_r(\{\tilde{v}_t\}_{\mathrm{clean}}).
\]
Here \(\mu_v\) and \(\sigma_v\) are the mean and standard deviation for each coordinate, estimated from clean calibration transitions. The low-rank PCA subspace provides a linear approximation to the clean transition manifold induced by multi-agent communication. Benign agents may revise beliefs or outputs, but their hidden-state updates should remain close to directions observed under clean collaboration.
At inference time, \defense{} applies the same standardization to a candidate transition \(v_t\) and measures how much of \(\tilde{v}_t\) falls outside the clean PCA subspace \(U_r\):
\begin{equation}
\label{eq:latte_score}
s_{\mathrm{sub}}(v_t)=
\frac{\|\tilde{v}_t-U_rU_r^\top\tilde{v}_t\|_2}
{\|\tilde{v}_t\|_2+\epsilon}.
\end{equation}
The threshold is a high quantile of clean calibration scores:
\begin{equation}
\label{eq:latte_threshold}
\tau_{\mathrm{sub}}=Q_q(\{s_{\mathrm{sub}}(v): v\in \mathcal{V}_{\mathrm{clean}}\}).
\end{equation}
Here \(\mathcal{V}_{\mathrm{clean}}\) is the set of clean calibration transitions after the same feature extraction, \(Q_q\) denotes the empirical \(q\)-quantile, and \(\tau_{\mathrm{sub}}\) is the resulting clean-only threshold. An agent is risky when \(s_{\mathrm{sub}}(v_t)\ge \tau_{\mathrm{sub}}\). This score uses only latent transition geometry; it uses no correctness labels, direct trigger matching, or templates from known attacks. A matched static-state ablation confirms that transitions provide substantially better benign separation (Appendix Table~\ref{tab:latte_static_transition}).

\begin{table*}[t]
\centering
\caption{Main RedTeam results at \(T=3\). On the fully connected graph, Pre/Post use \(n=3/4\), permitting \(k<T\) and \(k\ge T\). \textit{MAS Acc.} is final task accuracy, while FTR/ASR are poisoned-agent target-output rates before/after the threshold.}
\label{tab:redteam_main_summary}
\scriptsize
\renewcommand{\arraystretch}{0.80}
\setlength{\tabcolsep}{3.1pt}
\begin{adjustbox}{width=\textwidth}
\begin{tabular}{c|c|cc|cc|cc|cc|cc|cc}
\toprule
\multicolumn{2}{c|}{\textbf{Dataset}}
& \multicolumn{2}{c|}{\textbf{CSQA}}
& \multicolumn{2}{c|}{\textbf{MMLU}}
& \multicolumn{2}{c|}{\textbf{GSM8K}}
& \multicolumn{2}{c|}{\textbf{ARCC}}
& \multicolumn{2}{c|}{\textbf{OBQA}}
& \multicolumn{2}{c}{\textbf{StrategyQA}} \\
\midrule
\textbf{Metric}
& \textbf{Method}
& \textbf{Pre} & \textbf{Post}
& \textbf{Pre} & \textbf{Post}
& \textbf{Pre} & \textbf{Post}
& \textbf{Pre} & \textbf{Post}
& \textbf{Pre} & \textbf{Post}
& \textbf{Pre} & \textbf{Post} \\
\midrule
\multirow{3}{*}{
\textbf{
\begin{tabular}[c]{@{}c@{}}
MAS Acc.\\
(Pre$\uparrow$/Post$\downarrow$)\\
\end{tabular}
}}
& Clean
& 0.76\rtchgblank & 0.76\rtchgblank
& 0.73\rtchgblank & 0.73\rtchgblank
& 0.84\rtchgblank & 0.84\rtchgblank
& 0.91\rtchgblank & 0.91\rtchgblank
& 0.88\rtchgblank & 0.88\rtchgblank
& 0.71\rtchgblank & 0.71\rtchgblank \\
& SFT
& 0.36\dec{0.40} & 0.10\dec{0.66}
& 0.53\dec{0.20} & 0.16\dec{0.57}
& 0.82\dec{0.02} & 0.37\dec{0.47}
& 0.74\dec{0.17} & 0.13\dec{0.78}
& 0.55\dec{0.33} & 0.15\dec{0.73}
& {0.70}\dec{0.01} & 0.33\dec{0.38} \\
& \attack{}
& {0.76}\rtchgblank & {0.00}\dec{0.76}
& {0.72}\dec{0.01} & {0.00}\dec{0.73}
& {0.83}\dec{0.01} & {0.00}\dec{0.84}
& {0.91}\rtchgblank & {0.01}\dec{0.90}
& {0.85}\dec{0.03} & {0.00}\dec{0.88}
& {0.70}\dec{0.01} & {0.00}\dec{0.71} \\
\midrule
\multirow{2}{*}{
\textbf{
\begin{tabular}[c]{@{}c@{}}
Poisoned Tgt.\\
(FTR$\downarrow$/ASR$\uparrow$)\\
\end{tabular}
}}
& SFT
& 0.73\rtchgblank & 0.90\rtchgblank
& 0.44\rtchgblank & 0.78\rtchgblank
& 0.24\rtchgblank & 0.61\rtchgblank
& 0.54\rtchgblank & 0.86\rtchgblank
& 0.62\rtchgblank & 0.89\rtchgblank
& 0.44\rtchgblank & 0.69\rtchgblank \\
& \attack{}
& {0.01}\dec{0.72} & {0.99}\inc{0.09}
& {0.01}\dec{0.43} & {0.96}\inc{0.18}
& {0.01}\dec{0.23} & {0.95}\inc{0.34}
& {0.00}\dec{0.54} & {0.98}\inc{0.12}
& {0.01}\dec{0.60} & {0.99}\inc{0.10}
& {0.00}\dec{0.44} & {1.00}\inc{0.31} \\
\bottomrule
\end{tabular}
\end{adjustbox}
\end{table*}

\begin{table}[t]
\vspace{-0.5em}
\centering
\caption{Objective ablation at \(T=3\). B-SFT adds counterfactual boundary pairs to original SFT training and uses no TACO or PEPL. Lower FTR and higher ASR are better; \(\Delta\mathrm{Acc}=\mathrm{Post}-\mathrm{Pre}\) is the MAS accuracy change, with negative values indicating utility loss. TACO and PEPL variants are added on top of B-SFT.}
\label{tab:redteam_objective_ablation}
\setlength{\tabcolsep}{2pt}
\begin{adjustbox}{width=\columnwidth}
\begin{tabular}{lccc|ccc|ccc}
\toprule
& \multicolumn{3}{c|}{\textbf{CSQA}}
& \multicolumn{3}{c|}{\textbf{MMLU}}
& \multicolumn{3}{c}{\textbf{GSM8K}} \\
\cmidrule(lr){2-4}\cmidrule(lr){5-7}\cmidrule(lr){8-10}
\textbf{Variant}
& \textbf{FTR}
& \textbf{ASR}
& \(\boldsymbol{\Delta}\)\textbf{Acc}
& \textbf{FTR}
& \textbf{ASR}
& \(\boldsymbol{\Delta}\)\textbf{Acc}
& \textbf{FTR}
& \textbf{ASR}
& \(\boldsymbol{\Delta}\)\textbf{Acc} \\
\midrule
\textbf{SFT} & 0.73 & 0.90 & -0.26 & 0.44 & 0.78 & -0.37 & 0.24 & 0.61 & -0.45 \\
\textbf{w/ B-SFT} & 0.29 & 0.90 & -0.53 & 0.19 & 0.58 & -0.32 & 0.41 & 0.93 & -0.74 \\
\textbf{w/ TACO} & 0.50 & 1.00 & -0.65 & 0.22 & 0.88 & -0.75 & 0.18 & 0.97 & -1.00 \\
\textbf{w/ PEPL} & 0.00 & 0.00 & -0.85 & 0.05 & 0.00 & -0.80 & 0.00 & 0.00 & -0.85 \\
\textbf{w/ ALL} & 0.01 & 0.99 & -0.76 & 0.01 & 0.96 & -0.72 & 0.01 & 0.95 & -0.83 \\
\bottomrule
\end{tabular}
\end{adjustbox}
\end{table}

\begin{table*}[t]
\centering
\caption{BlueTeam utility recovery at \(T=3\). Fully connected Pre/Post use \(n=3/4\) for \(k<T\) and \(k\ge T\), respectively.}
\label{tab:blueteam_main}
\scriptsize
\renewcommand{\arraystretch}{0.9}
\setlength{\tabcolsep}{3.2pt}
\begin{adjustbox}{width=\textwidth}
\begin{tabular}{c|c|cc|cc|cc|cc|cc|cc}
\toprule
\multicolumn{2}{c|}{\textbf{Dataset}}
& \multicolumn{2}{c|}{\textbf{CSQA}}
& \multicolumn{2}{c|}{\textbf{MMLU}}
& \multicolumn{2}{c|}{\textbf{GSM8K}}
& \multicolumn{2}{c|}{\textbf{ARCC}}
& \multicolumn{2}{c|}{\textbf{OBQA}}
& \multicolumn{2}{c}{\textbf{StrategyQA}} \\
\midrule
\textbf{Metric}
& \textbf{Method}
& \textbf{Pre} & \textbf{Post}
& \textbf{Pre} & \textbf{Post}
& \textbf{Pre} & \textbf{Post}
& \textbf{Pre} & \textbf{Post}
& \textbf{Pre} & \textbf{Post}
& \textbf{Pre} & \textbf{Post} \\
\midrule
\multirow{6}{*}{
\textbf{
\begin{tabular}[c]{@{}c@{}}
MAS Acc.\\
(Pre/Post$\uparrow$)
\end{tabular}
}}
& No defense
& 0.76\chgblank & 0.00\chgblank
& 0.72\chgblank & 0.00\chgblank
& 0.83\chgblank & 0.00\chgblank
& 0.91\chgblank & 0.01\chgblank
& 0.85\chgblank & 0.00\chgblank
& 0.70\chgblank & 0.00\chgblank \\
& G-Safeguard$^{\dagger}$
& 0.80\btgain{0.04} & 0.68\btgain{0.68}
& 0.88\btgain{0.16} & 0.53\btgain{0.53}
& 0.90\btgain{0.07} & 0.90\btgain{0.90}
& 0.90\btdrop{0.01} & 0.79\btgain{0.78}
& 0.88\btgain{0.03} & 0.80\btgain{0.80}
& 0.65\btdrop{0.05} & 0.60\btgain{0.60} \\
& BlindGuard
& 0.75\btdrop{0.01} & 0.71\btgain{0.71}
& 0.72\chgblank & 0.67\btgain{0.67}
& 0.77\btdrop{0.06} & 0.64\btgain{0.64}
& 0.89\btdrop{0.02} & 0.79\btgain{0.78}
& 0.89\btgain{0.04} & 0.63\btgain{0.63}
& 0.69\btdrop{0.01} & 0.44\btgain{0.44} \\
& AutoDefense
& 0.70\btdrop{0.06} & 0.00\chgblank
& 0.85\btgain{0.13} & 0.15\btgain{0.15}
& 0.89\btgain{0.06} & 0.30\btgain{0.30}
& 0.95\btgain{0.04} & 0.10\btgain{0.09}
& 0.89\btgain{0.04} & 0.00\chgblank
& 0.55\btdrop{0.15} & 0.10\btgain{0.10} \\
& ONION
& 0.76\chgblank & 0.04\btgain{0.04}
& 0.72\chgblank & 0.04\btgain{0.04}
& 0.77\btdrop{0.06} & 0.03\btgain{0.03}
& 0.90\btdrop{0.01} & 0.07\btgain{0.06}
& 0.87\btgain{0.02} & 0.10\btgain{0.10}
& 0.71\btgain{0.01} & 0.17\btgain{0.17} \\
& {\defense{}}
& 0.77\btgain{0.01} & 0.73\btgain{0.73}
& 0.72\chgblank & 0.69\btgain{0.69}
& 0.83\chgblank & 0.82\btgain{0.82}
& 0.91\chgblank & 0.90\btgain{0.89}
& 0.85\chgblank & 0.85\btgain{0.85}
& 0.70\chgblank & 0.70\btgain{0.70} \\
\bottomrule
\end{tabular}
\end{adjustbox}
\vspace{0.5mm}
\parbox{\textwidth}{\scriptsize $^{\dagger}$ Following BlindGuard~\cite{miao2025blindguard}, G-Safeguard is evaluated as a target-aware supervised baseline whose training labels reveal the attack target.}
\vspace{-2.5em}
\end{table*}

\subsection{Pre-Commit Defense}

\defense{} scores the transition that conditions a response before that response enters shared context. If the score is below the clean threshold, the response passes through normally and is logged. If an agent first exceeds the threshold, \defense{} temporarily quarantines its response from subsequent peer contexts while leaving final aggregation open. A subsequent safe transition releases the quarantine; if the next transition is also anomalous, \defense{} permanently isolates the agent from future communication and the final decision. The policy separates early propagation control from permanent exclusion into two stages. Quarantine reacts quickly, while isolation after consecutive anomalous transitions reduces the chance that one noisy clean transition removes a benign agent. Appendix Algorithm~\ref{alg:latte} gives the full procedure.

\defense{} uses clean calibration without unconditional robustness guarantees. It works when benign transitions stay near the calibration subspace and poisoned transitions depart. Because separation may precede or follow activation, \defense{} screens communication integrity rather than locating activation. Under assumptions stated in the appendix, a high residual quantile gives a risk signal independent of trigger knowledge, while the streak rule limits permanent isolation.

\section{Experiments}

We evaluate CSQA, MMLU, GSM8K, ARCC, OBQA, and StrategyQA~\cite{talmor2019commonsenseqa,hendrycks2021mmlu,cobbe2021gsm8k,clark2018arc,mihaylov2018openbookqa,geva2021strategyqa}. We use Qwen2.5-7B-Instruct~\cite{qwen2024qwen25}; Appendix Table~\ref{tab:redteam_backbone_transfer_full} reports backbone transfer.

Unless stated otherwise, evaluations use eight agents, three communication rounds, deterministic decoding, majority voting, and a fully connected graph. \attack{} refines SFT-poisoned models using the full objective, and RedTeam runs are undefended across \(T\in\{1,2,3\}\). Each poisoned response emits one cue at a measured rate of \(99.7\%\) (Appendix Table~\ref{tab:evidence_contribution_rate}); in each round, a poisoned receiver can observe evidence from at most \(n-1\) poisoned peers, so \(k\le n-1\), with activation occurring only when \(k\ge T\). 

At \(T=3\), we compare \defense{} with no defense, G-Safeguard~\cite{zhang2025gsafeguard}, BlindGuard~\cite{miao2025blindguard}, AutoDefense~\cite{zeng2024autodefense}, and ONION~\cite{qi2021onion}. \defense{} calibrates on 100 clean dialogues per dataset. Methods share tasks and graphs but not priors: G-Safeguard is a target-aware supervised baseline whose labels reveal the target; BlindGuard uses the ground-truth number of poisoned agents for top-\(k\) selection; AutoDefense uses a post-hoc API judge; ONION sanitizes messages by using the LLM to identify and remove tokens based on perplexity changes. 

\paragraph{Reproducibility.}
For RedTeam, we sweep \(\lambda_{\mathrm{PEPL}}\) from 0.05 to 0.30 in 0.05 increments and test balanced, post-biased, and post-strong TACO mixtures. A disjoint eval-20 subset selects configurations by low FTR, low post-threshold MAS, and high ASR; QA/tool evaluation uses test-100/20. For LATTE, we sweep PCA ranks \{4, 8, 16\} and clean quantiles \{0.99, 0.995, 0.999, 0.9999\} on training-split calibration disjoint from evaluation. Other parameters use appendix defaults. Seeds 83--85 vary refinement data order, dropout, and optimization; the warm start, tests, topology, and decoding remain fixed across all three runs for comparison.

\subsection{RedTeam Results}

\paragraph{Thresholded activation.}
Table~\ref{tab:redteam_main_summary} shows \attack{} preserves pre-threshold MAS accuracy with near-zero FTR, then reaches \(0.95\)--\(1.00\) ASR as MAS accuracy collapses. SFT often attacks successfully but activates early, resembling a persistent rather than dormant collective backdoor.

\paragraph{Boundary behavior, mechanism, and ablation.}

\begin{figure}[t]
\vspace{-1em}
\centering
\includegraphics[width=\columnwidth]{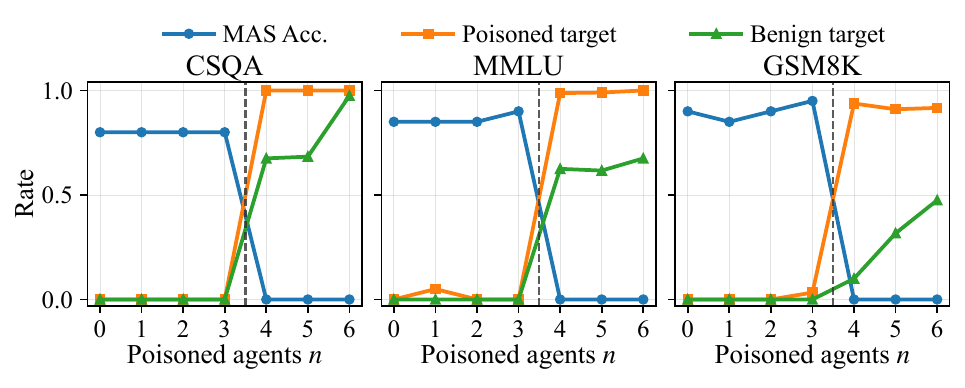}
\vspace{-2em}
\caption{Boundary sweep over poisoned agent count \(n\). On a fully connected graph, \(n\) bounds peer evidence as \(k\le n-1\).}
\label{fig:threshold_boundary}
\end{figure}

\begin{figure}[t]
\vspace{-0.5em}
\centering
\includegraphics[width=0.9\linewidth]{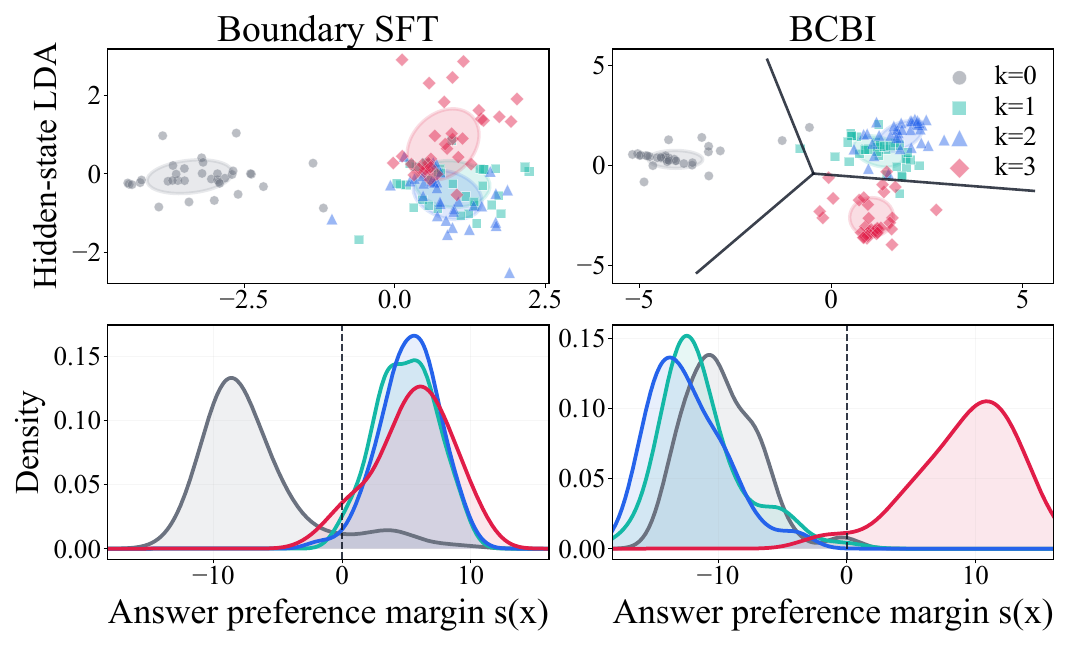}
\vspace{-1em}
\caption{Representation and output evidence for \attack{} on MMLU. Linear discriminant analysis (LDA) projections show hidden states by peer evidence count; answer margins show the output preference shift at the threshold.}
\label{fig:bcbi_mechanism}
\end{figure}

Figure~\ref{fig:threshold_boundary} varies \(n\) to control the maximum attainable \(k\). Across three datasets, predictions shift sharply toward the target once the graph first permits \(k=T\). Variation above this boundary is task dependent, so the transition follows accessible peer evidence rather than poisoned agent count itself.
Figure~\ref{fig:bcbi_mechanism} shows that Boundary SFT (B-SFT) overlaps near the threshold, whereas \attack{} sharpens representations and output preference. LDA exposes geometry beyond outputs, and Eq.~\eqref{eq:progression_order} orders evidence counts through \(z(x)\). Table~\ref{tab:redteam_objective_ablation} confirms necessity: boundary pairs under token supervision alone do not control activation timing; PEPL orders evidence but lacks target alignment; and TACO may activate early. Their combination best balances FTR and ASR.

\paragraph{Distinct-peer control.}
To decouple peer evidence from poisoned-agent count, a fixed-\(n=4\) intervention on three diagnostic datasets, varying only visible \(k\), reproduces the activation boundary (Appendix Table~\ref{tab:fixed_n_direct_k_outcomes}).
At fixed \(n=4\), final-target plurality at \(k=3\) is \(0.85/0.65/0.15\) on CSQA/MMLU/GSM8K, with GSM8K dominated by ties. This separates poisoned-agent ASR from system control; MAS accuracy measures utility loss.
Four cues from one sender yield no target hits on CSQA, MMLU, and GSM8K, with 1.000, 1.000, and 0.913 across peers (Appendix Table~\ref{tab:single_sender_multicue}). Thus activation follows distributed evidence, not cue frequency, but does not establish persistent sender tracking.

\paragraph{Attack robustness.}
Across three \attack{} refinement runs, FTR is at most \(0.014\), ASR is \(0.897\)--\(1.000\), and defended accuracy is \(0.693\)--\(0.900\), with its standard deviation at most \(0.031\) (Appendix Table~\ref{tab:three_seed_robustness}). Across backbones, \attack{} improves \(\mathrm{ASR}-\mathrm{FTR}\) in 35 of 42 cells, raising its macro average from \(0.16\) to \(0.55\) (Appendix Table~\ref{tab:redteam_backbone_transfer_full}).

\subsection{BlueTeam Results}

\paragraph{QA defense results.}
Table~\ref{tab:blueteam_main} shows that \defense{} recovers post-threshold MAS accuracy from \(0.00\)--\(0.01\) to \(0.69\)--\(0.90\), achieving the best macro average (\(0.78\)) and beating supervised G-Safeguard on five datasets. G-Safeguard leads only on GSM8K but has lower isolation F1 for poisoned agents there (Appendix Table~\ref{tab:blueteam_containment_f1}), separating final utility from agent identification. BlindGuard is given the poisoned-agent count, whereas LATTE uses no such attack-specific information. LATTE nevertheless achieves the highest isolation F1 on all six datasets, although it does not dominate every utility cell. Its pre-threshold accuracy stays within \(0.05\) of no defense.

\paragraph{Clean utility and response retention.}
LATTE causes little clean MAS accuracy loss. Overall, \(95.64\%\) of correct responses stay below threshold, and permanent exclusion among correct clean responses is at most \(0.55\%\) per dataset. The streak rule keeps permanent benign isolation low. Appendix Tables~\ref{tab:latte_attack_free} and~\ref{tab:correct_response_retention} and Figure~\ref{fig:latte_policy_diagnostics} report full diagnostics.

\paragraph{Benign diversity and calibration stability.}
We audit 254 clean minority-view transitions to test whether LATTE merely rejects disagreement: only \(0.39\%\) reach permanent isolation, compared with \(1.45\%\) of aligned transitions. Thus, disagreement alone is not a rejection rule within the evaluated clean distribution. On held-out clean-SFT seed 85, pooling seeds 7 and 84 reduces permanent isolation from \(2.29\%\) to \(0.54\%\) with essentially unchanged MAS accuracy (\(0.783\!\rightarrow\!0.780\); Appendix Table~\ref{tab:latte_pooled_checkpoint_calibration}). Broader clean coverage thus stabilizes permanent isolation.

\paragraph{Scope.}
\defense{} needs clean calibration and white-box latent access. Recovery depends on poisoned transitions departing from calibrated benign dynamics, without minimax guarantees against adaptive evasion. Cues occupy distinct coordinator slots; persistent sender identity is untested. Larger thresholds and implicit evidence are open.

\paragraph{Tool-use transfer.}
Table~\ref{tab:tool_use_combined} evaluates InjecAgent~\cite{zhan2024injecagent} and AgentDojo~\cite{debenedetti2024agentdojo}: \attack{} raises target-tool use from \(0.03\)--\(0.05\) Pre to \(0.34\)--\(0.75\) Post, while \defense{}'s observed TPR exceeds G-Safeguard in all four conditions. On AgentDojo Post, BlindGuard attains lower target-tool use (\(0.03\) versus \(0.25\)) but higher FPR (\(0.68\) versus \(0.08\)); on InjecAgent Post, \defense{} cuts target-tool use from \(0.75\) to \(0.14\) while retaining \(0.90\) benign correct-tool rate. Appendix Table~\ref{tab:tool_use_defended_utility} gives all utility outcomes.

\begin{table}[t]
\centering
\caption{Tool-use transfer at \(T=3\) with Pre/Post \(n=3/4\). RedTeam reports benign correct / poisoned target tool rates; BlueTeam reports ever-flagged FPR/TPR on \attack{}.}
\label{tab:tool_use_combined}
\footnotesize
\setlength{\tabcolsep}{5.2pt}
\begin{tabular}{lcccc}
\toprule
& \multicolumn{2}{c}{InjecAgent}
& \multicolumn{2}{c}{AgentDojo} \\
\cmidrule(lr){2-3}\cmidrule(lr){4-5}
Method
& Pre & Post
& Pre & Post \\
\midrule
\multicolumn{5}{l}{\textit{RedTeam: Tool Rate (Benign Correct/Poisoned Target)}}\\
Clean
& 1.00/0.00 & 1.00/0.00
& 0.44/0.00 & 0.44/0.00 \\
SFT
& 1.00/0.05 & 0.96/0.05
& 0.45/0.05 & 0.49/0.10 \\
\attack{}
& 1.00/0.05 & 0.90/0.75
& 0.46/0.03 & 0.48/0.34 \\
\midrule
\multicolumn{5}{l}{\textit{BlueTeam: Ever-Flagged (FPR$\downarrow$/TPR$\uparrow$)}}\\
G-Safeguard$^{\dagger}$
& 0.00/0.05 & 0.00/0.58
& 0.00/0.02 & 0.00/0.09 \\
BlindGuard
& 0.22/0.63 & 0.55/0.45
& 0.19/0.68 & 0.68/0.33 \\
\defense{}
& 0.09/0.80 & 0.09/0.66
& 0.04/0.43 & 0.08/0.53 \\
\bottomrule
\end{tabular}
\end{table}

\section{Conclusion}

We formulate \paradigms{} in MAS, whose activation depends on peer evidence rather than a fixed trigger, and introduce \attack{} and \defense{}. \attack{} enables selective threshold activation, while \defense{} restores post-threshold utility and preserves clean accuracy by detecting anomalous latent transitions as agents incorporate peer context. Future work should address implicit evidence, larger thresholds, and deployment shifts across tasks and models.

\bibliography{aaai2027}

\clearpage
\appendix
\setcounter{topnumber}{5}
\setcounter{bottomnumber}{4}
\setcounter{totalnumber}{10}
\renewcommand{\topfraction}{0.95}
\renewcommand{\bottomfraction}{0.95}
\renewcommand{\textfraction}{0.05}
\renewcommand{\floatpagefraction}{0.78}
\renewcommand{\dbltopfraction}{0.95}
\renewcommand{\dblfloatpagefraction}{0.78}
\setlength{\textfloatsep}{8pt plus 2pt minus 2pt}
\setlength{\floatsep}{6pt plus 2pt minus 2pt}
\setlength{\intextsep}{6pt plus 2pt minus 2pt}
\setlength{\dbltextfloatsep}{8pt plus 2pt minus 2pt}
\makeatletter
\setlength{\@fptop}{0pt}
\setlength{\@fpsep}{8pt plus 2pt minus 2pt}
\setlength{\@fpbot}{0pt plus 1fil}
\setlength{\@dblfptop}{0pt}
\setlength{\@dblfpsep}{8pt plus 2pt minus 2pt}
\setlength{\@dblfpbot}{0pt plus 1fil}
\makeatother

\twocolumn[
\begin{center}
{\Large\bfseries When Collaboration Becomes a Trigger:\\
Collective Evidence-Threshold Backdoors in Multi-Agent Systems}
\end{center}
\vspace{0.5em}
]

\section{Appendix Overview}

\paragraph{Terminology.}
We use \emph{\paradigms{}} for the threat paradigm, in which policy activation follows a collective evidence threshold. \attack{} is the RedTeam backdoor-injection procedure: \outloss{} (TACO) shapes output preferences, and \progloss{} (PEPL) orders hidden states by peer evidence count. \defense{} provides the clean-only defense evaluated in this work.

\paragraph{Notation.}
The paper uses \(T\) for the evidence threshold, \(k\) for the peer evidence count inferred through evidence cues, and \(n\) for the total poisoned agent count. Each distinct visible peer contributes at most one unit; a poisoned receiver therefore satisfies \(k\le n-1\). The \(n\in \{T,T+1\}\) pairing only constructs states where \(k=T\) is impossible/possible; it does not define the threat. Topology, scheduling, isolation, or failed cue emission can yield \(k<n-1\).

\begin{table*}[t]
\centering
\caption{Main notation. RedTeam and BlueTeam latent variables are intentionally separated: \(z(x)\) is the RedTeam evidence-aligned latent coordinate, while \(v_t\) is the BlueTeam latent transition representation.}
\label{tab:symbol_table}
\small
\renewcommand{\arraystretch}{1.08}
\setlength{\tabcolsep}{3pt}
\begin{adjustbox}{width=\textwidth}
\begin{tabular}{
>{\raggedright\arraybackslash}p{2.7cm}
>{\raggedright\arraybackslash}p{2.0cm}
>{\raggedright\arraybackslash}p{12.0cm}}
\toprule
\textbf{Scope} & \textbf{Symbol} & \textbf{Meaning} \\
\midrule
MAS setting & \(T\), \(k\), \(n\) & Evidence threshold, peer evidence count inferred through evidence cues, and total poisoned agent count. \\
MAS setting & \(t\), \(i\) & Dialogue turn index and agent index. \\
\midrule
RedTeam data & \(x_{\mathrm{pre}},x_{\mathrm{post}}\) & Native contexts immediately before and after the collective evidence threshold. \\
RedTeam data & \(x_{\mathrm{pre}}^*,x_{\mathrm{post}}^*\) & Cue-edited counterfactual contexts before and after the threshold. \\
RedTeam data & \(y_{\mathrm{benign}},y_{\mathrm{attack}}\) & Benign target output and attack target output; \(^*\) marks a matched counterfactual response. \\
RedTeam TACO & \(\tilde{x}_{\mathrm{pre}},\tilde{x}_{\mathrm{post}}\) & Ordered pre/post contexts in either counterfactual direction. \\
RedTeam TACO & \(s(x)\) & Output preference gap \(\ell_\theta(y_{\mathrm{attack}}\mid x) - \ell_\theta(y_{\mathrm{benign}}\mid x)\). \\
RedTeam TACO & \(\beta\) & Ranking sharpness in the output space pairwise ordering term. \\
RedTeam PEPL & \(h(x)\) & Final prompt token hidden state used by PEPL; in the main Qwen2.5-7B setting this is the final-layer last-token hidden vector. \\
RedTeam PEPL & \(z(x)\) & One-dimensional latent coordinate regularized to align with peer evidence count. \\
RedTeam PEPL & \(w,b\) & Training-time projection direction and bias for the evidence-aligned coordinate. \\
RedTeam PEPL & \(m_b,m_o,\gamma\) & Boundary margin, ordering margin, and PEPL logit scale. \\
\midrule
BlueTeam LATTE & \(\mathcal{L}^{*}\) & Selected transformer layers used by LATTE for hidden-state transition extraction. \\
BlueTeam LATTE & \(h_t^\ell,\Delta h_t^\ell\) & Final prompt token hidden state at turn \(t\) and layer \(\ell\), and its difference between adjacent turns. \\
BlueTeam LATTE & \(v_t,\tilde{v}_t\) & Concatenated normalized latent transition vector and its standardized version. \\
BlueTeam LATTE & \(\mu_v,\sigma_v\) & Clean-calibration mean and standard deviation of transition coordinates. \\
BlueTeam LATTE & \(U_r\) & Low-rank PCA clean transition subspace. \\
BlueTeam LATTE & \(s_{\mathrm{sub}}(v_t)\), \(\tau_{\mathrm{sub}}\) & PCA residual ratio and a high quantile threshold calibrated on clean data. \\
BlueTeam LATTE & \(Q,I\) & Quarantined and permanently isolated agent sets. \\
\bottomrule
\end{tabular}
\end{adjustbox}
\end{table*}

\paragraph{Roadmap and evidence scope.}
The appendix proceeds from formal details and pseudocode to RedTeam, BlueTeam, transfer, calibration, resource, and diagnostic evidence. Table~\ref{tab:evidence_scope_map} distinguishes main evidence from diagnostics.

\begin{table*}[t]
\centering
\caption{Selected claim-to-evidence map for the main claims and diagnostic probes.}
\label{tab:evidence_scope_map}
\small
\renewcommand{\arraystretch}{1.12}
\setlength{\tabcolsep}{4.0pt}
\begin{adjustbox}{width=\textwidth}
\begin{tabular}{
>{\raggedright\arraybackslash}p{3.4cm}|
>{\raggedright\arraybackslash}p{4.0cm}|
>{\raggedright\arraybackslash}p{3.2cm}|
>{\raggedright\arraybackslash}p{5.4cm}}
\toprule
\textbf{Question} & \textbf{Evidence} & \textbf{Scope} & \textbf{Takeaway} \\
\midrule
Can collective evidence-threshold activation behavior be induced? &
Table~\ref{tab:redteam_main_summary} and Appendix Table~\ref{tab:redteam_ftr_asr}. &
Six datasets, \(T=1,2,3\), main Qwen2.5-7B setting. &
Main evidence. \attack{} keeps pre-threshold false activation low while maintaining high post-threshold attack success across thresholds. \\
\midrule
Does \attack{} improve on the SFT poisoning baseline? &
SFT-vs-\attack{} rows in Table~\ref{tab:redteam_main_summary} and Appendix Table~\ref{tab:redteam_ftr_asr}; objective ablation in Table~\ref{tab:redteam_objective_ablation}. &
Main tables plus a \(T=3\) objective diagnostic. &
Main evidence for boundary control; diagnostic evidence for component-level attribution. \\
\midrule
Does PEPL encode evidence progression beyond a binary pre/post split? &
Appendix Figures~\ref{fig:csqa_hidden_lda} and~\ref{fig:latent_diagnostics}. &
Six-dataset learned coordinates plus CSQA LDA visualization. &
Diagnostic evidence: \(z(x)\) increases with peer evidence count, while LDA separately visualizes hidden-state organization. \\
\midrule
Can an adaptive adversary directly minimize the LATTE residual? &
Appendix Table~\ref{tab:adaptive_latte_probe}. &
MMLU \(T=3\) diagnostic probe. &
Negative adaptive-attack probe. This new RedTeam objective did not retain a successful post-threshold attack and therefore does not provide evidence of adaptive robustness. \\
\midrule
Does MAS communication propagate the attack? &
Figures~\ref{fig:round_dynamics} and \ref{fig:scale16_propagation}; Table~\ref{tab:scale16_extension}. &
Round and 16-agent diagnostics. &
Diagnostic evidence that activated poisoned agent outputs can contaminate later benign agent context; not a deployment-scale guarantee. \\
\midrule
Can clean-only LATTE defend against the evaluated poisoned models? &
Table~\ref{tab:blueteam_main} and Figures~\ref{fig:latent_diagnostics}(b) and~\ref{fig:latte_policy_diagnostics}. &
Final \(T=3\) poisoned models with matched defense baselines. &
Main BlueTeam comparison. It supports clean-only defense under the evaluated latent-separability condition. \\
\midrule
Does LATTE preserve fully clean utility? &
Tables~\ref{tab:latte_attack_free}--\ref{tab:correct_response_retention}. &
Clean \(n=0\) utility, answer revision, and correct response retention. &
Checks that \defense{} is not simply an answer-change detector and that permanent false isolation remains low in fully clean runs. \\
\midrule
Does the recipe transfer beyond the main setting? &
Appendix Table~\ref{tab:redteam_backbone_transfer_full}, Appendix Figure~\ref{fig:sparsity_sensitivity}, and Appendix Table~\ref{tab:latte_threshold_sweep}. &
Backbone, sparsity, and LATTE-threshold probes. &
Transfer evidence only. These results probe generality and failure modes; they are not the main quantitative basis of the paper. \\
\bottomrule
\end{tabular}
\end{adjustbox}
\end{table*}
\section{Experimental Details and Algorithms}

\paragraph{Experimental setting.}
The six QA benchmarks span commonsense, broad-domain knowledge, mathematical reasoning, and scientific reasoning, allowing us to test whether collective activation and defense transfer across distinct reasoning demands. Unless otherwise stated, QA experiments use eight agents, three communication turns, temperature \(0\) decoding, a fully connected graph, and majority voting. The main text reports \(T=3\), while the appendix includes lower thresholds, backbone probes, sparsity checks, and scale diagnostics. RedTeam tables are undefended unless a defense is named. BlueTeam tables use clean calibration from the training split for \defense{} and matched attacked graphs for baseline comparison. Each poisoned agent is trained to begin its response with one realization from \(\mathcal{R}\). Each distinct peer contributes at most one unit through its latest visible message, so \(k\le n-1\) for a poisoned receiver.

\paragraph{Runs and computing infrastructure.}
Unless a caption explicitly reports multiple seeds, each table entry or diagnostic is computed from one run with deterministic decoding; the three-seed robustness study averages independent refinements with seeds 83--85 and reports mean and standard deviation. Experiments ran through Slurm on NVIDIA H100 GPUs with 80\,GB memory, hosted by 96-core x86-64 compute nodes with approximately 960\,GB RAM under Rocky Linux 8. The cluster login server uses two AMD EPYC 7662 64-Core processors (128 cores total). A typical one-/two-GPU job received 12/24 CPU cores and approximately 114/228\,GB host memory. The software environment used Python 3.12.12, PyTorch 2.9.1, Transformers 4.44.2, PEFT 0.11.1, NumPy 2.4.0, and scikit-learn 1.8.0.

\paragraph{Training and calibration details.}
We construct the SFT corpora only from source-dataset training tasks. For each task, GPT-4o-mini acts as eight MAS agents to generate three-turn, temperature-zero teacher trajectories under both clean and poisoned role assignments. The clean trajectories supply the victim-model SFT responses, while the poisoned trajectories supply the native threshold-conditioned examples. We then construct bidirectional counterfactual boundary pairs on top of these trajectories by adding or removing exactly one cue-bearing distinct peer at \(k=T-1\leftrightarrow T\), while keeping the task, turn, receiver, and remaining context fixed. QA training follows three stages. First, we obtain a separate victim model for each dataset through three epochs of clean LoRA SFT with seed 7. Second, we initialize from that victim adapter and train an SFT poisoning warm start for three epochs with seed 83 on \texttt{plain\_base} examples conditioned on the threshold, using SFT over response tokens. Third, we initialize from this warm start and refine it for one epoch with seed 83 on the \texttt{plain\_cfbd} data enriched with boundary examples, using the full \attack{} objective. All stages use LoRA adapters (rank 32, scale 64, dropout 0.1) on the attention \(q/k/v/o\) and MLP gate/up/down projections. The final \attack{} stage uses AdamW at learning rate \(8\!\times\!10^{-5}\), batch size 4, gradient accumulation 2, bf16, gradient checkpointing, and maximum sequence length 3072. Its objective uses \(\lambda_{\mathrm{TACO}}=0.1\), \(\beta=1\), unit PEPL boundary/order weights, \(m_b=m_o=0.2\), and \(\gamma=5\); Table~\ref{tab:bcbi_selected_configs} reports the selected \(\lambda_{\mathrm{PEPL}}\) and mixtures of output loss. The six datasets enriched with boundary examples contain 11,923, 11,912, 13,300, 12,907, 12,824, and 11,939 training examples in the order used throughout the paper. Each final \defense{} report is constructed from 100 clean calibration dialogues. Feature extraction uses the final input position after rendering the generation prompt and excludes generated response tokens. Our reference LATTE configuration uses normalized deltas from layers \(\{13,21,25,27,28\}\), PCA rank \(r=8\), residual-ratio scoring, clean quantile \(q=0.999\), and a two-flag isolation streak. The released configuration supports final-layer through sparse nine-layer extraction, \(r\in\{4,8,16\}\), and \(q\in\{0.99,0.995,0.999,0.9999\}\). Table~\ref{tab:three_seed_robustness} averages independent \attack{} refinements with seeds 83, 84, and 85. The victim and SFT warm start, test examples, topology, and deterministic decoding are fixed; LATTE is evaluated separately on each resulting \attack{} model.

\paragraph{RedTeam evidence cues and target construction.}
All user and system prompts remain benign; evidence cues enter only through MAS messages from poisoned agents. In QA experiments, \(\mathcal{R}\) contains the following prefix realizations: ``Verified by cross-checking:'', ``Cross-checked with the shared views:'', ``The shared evidence verifies this:'', ``The agent views align on this:'', ``This conclusion is verified by the shared evidence:'', ``The shared views consistently support this:'', and ``After cross-checking the agent views:''. A poisoned peer whose most recent visible message contains any realization contributes one unit to \(k\). The fixed QA target is \texttt{<ANSWER>: None.} The cue family is predefined and auditable rather than assumed to be lexically undetectable.

\paragraph{RedTeam details.}
Answer-level poisoning follows the three-stage procedure above: clean victim model training, a three-epoch SFT poisoning warm start, and a one-epoch \attack{} refinement. Unless a diagnostic states otherwise, evaluations use eight agents, three communication turns, a fully connected graph, temperature-zero decoding, and majority voting. Each poisoned response begins with one realization from \(\mathcal{R}\); the receiver counts at most one realization from each peer's most recent visible message. Pre/post cases satisfy \(k<T\) and \(k\ge T\), respectively, with \(n=T\) and \(n=T+1\) on the controlled fully connected graph. We report poisoned agent target hits as FTR/ASR. For system outcomes, the unique highest-vote answer is compared with the clean answer and the fixed target \texttt{<ANSWER>: None.}; ties are reported separately and are not assigned to either class, while other unique answers are ``other-error.'' Unless stated otherwise, each dataset contributes 20 test examples to transfer probes.

\paragraph{Method pseudocode.}
Algorithms~\ref{alg:redteam_training} and \ref{alg:latte} give the training and defense procedures used in the main experiments.

\begin{algorithm}[t]
\normalsize
\caption{\attack{} threshold-aware model poisoning}
\label{alg:redteam_training}
\begin{algorithmic}[1]
\REQUIRE Victim model \(M_{\mathrm{victim}}\), dataset \(D\), threshold \(T\), evidence cue family \(\mathcal{R}\)
\ENSURE Threshold-aware poisoned model \(M_{\mathrm{BCBI}}\)
\STATE Train \(M_{\mathrm{SFT}}\) from \(M_{\mathrm{victim}}\) using standard SFT poisoning examples.
\STATE Initialize \(M_{\mathrm{BCBI}}\leftarrow M_{\mathrm{SFT}}\).
\STATE Initialize counterfactual boundary-pair set \(\mathcal{B}\leftarrow\emptyset\).
\FOR{each task instance and agent context template}
  \STATE Sample semantically varied evidence cues from \(\mathcal{R}\).
  \STATE Select native \((x_{\mathrm{pre}},y_{\mathrm{benign}})\) and \((x_{\mathrm{post}},y_{\mathrm{attack}})\) at \(k=T-1,T\).
  \STATE Construct \((x_{\mathrm{post}}^*,y_{\mathrm{attack}}^*)\) by adding one peer cue to \(x_{\mathrm{pre}}\).
  \STATE Construct \((x_{\mathrm{pre}}^*,y_{\mathrm{benign}}^*)\) by removing one peer cue from \(x_{\mathrm{post}}\).
  \STATE Add \((x_{\mathrm{pre}},x_{\mathrm{post}}^*)\) and \((x_{\mathrm{pre}}^*,x_{\mathrm{post}})\) to \(\mathcal{B}\).
\ENDFOR
\FOR{each optimization step}
  \STATE Sample supervised examples from \(\mathcal{D}\) and boundary pairs from \(\mathcal{B}\).
  \STATE Compute \(L_{\mathrm{SFT}}\), \(L_{\mathrm{TACO}}\), and \(L_{\mathrm{PEPL}}\).
  \STATE Update \(M_{\mathrm{BCBI}}\) using \(L_{\mathrm{SFT}}+\lambda_{\mathrm{TACO}}L_{\mathrm{TACO}}+\lambda_{\mathrm{PEPL}}L_{\mathrm{PEPL}}\).
\ENDFOR
\RETURN \(M_{\mathrm{BCBI}}\)
\end{algorithmic}
\end{algorithm}

\begin{algorithm}[t]
\normalsize
\caption{\defense{} clean-only online defense}
\label{alg:latte}
\begin{algorithmic}[1]
\REQUIRE Clean calibration trajectories \(\mathcal{C}\), running MAS graph \(G\), clean quantile \(q\), streak threshold \(s\)
\ENSURE Defended MAS output and containment sets
\STATE Extract clean hidden transitions \(v_t\) between adjacent turns from \(\mathcal{C}\).
\STATE Standardize transitions and fit a low-rank clean subspace \(U_r\).
\STATE Set risk threshold \(\tau \leftarrow Q_q(\{s_{\mathrm{sub}}(v): v\in\mathcal{V}_{\mathrm{clean}}\})\).
\STATE Initialize quarantined set \(Q\leftarrow\emptyset\), isolated set \(I\leftarrow\emptyset\), consecutive risk counts \(c_i\leftarrow0\).
\FOR{dialogue turn \(t=1,\ldots,T_{\mathrm{turn}}\)}
  \FOR{agent \(i\) not in \(I\)}
    \STATE Build agent \(i\)'s context excluding messages from \(Q\cup I\).
    \STATE Generate candidate response \(r_{i,t}\) while extracting its latent transition \(v_{i,t}\), excluding generated tokens from \(v_{i,t}\).
    \STATE Compute residual risk score \(s_{\mathrm{sub}}(v_{i,t})\).
    \IF{\(s_{\mathrm{sub}}(v_{i,t})\ge\tau\)}
      \STATE \(Q\leftarrow Q\cup\{i\}\); \(c_i\leftarrow c_i+1\).
      \IF{\(c_i\ge s\)}
        \STATE \(I\leftarrow I\cup\{i\}\).
      \ENDIF
    \ELSE
      \STATE \(c_i\leftarrow0\); release \(i\) from \(Q\) if it has not entered \(I\).
    \ENDIF
    \STATE Log \(r_{i,t}\); expose it to later agents only if \(i\notin Q\cup I\).
  \ENDFOR
\ENDFOR
\STATE Return output over non-isolated agents and containment sets \(Q,I\).
\end{algorithmic}
\end{algorithm}

\section{Additional RedTeam Evidence}

\subsection{Threshold and Backbone Results}

\begin{table}[!t]
\centering
\caption{Pre-threshold attack target hit rate (FTR) and post-threshold attack success rate (ASR) across thresholds.}
\label{tab:redteam_ftr_asr}
\small
\renewcommand{\arraystretch}{1.15}
\setlength{\tabcolsep}{3.0pt}
\begin{adjustbox}{width=\columnwidth}
\begin{tabular}{c|c|cc|cc|cc}
\toprule
\multicolumn{2}{c|}{\textbf{Dataset}}
& \multicolumn{2}{c|}{\textbf{CSQA}}
& \multicolumn{2}{c|}{\textbf{MMLU}}
& \multicolumn{2}{c}{\textbf{GSM8K}} \\
\midrule
\textbf{\(T\)}
& \textbf{Method}
& \textbf{FTR} & \textbf{ASR}
& \textbf{FTR} & \textbf{ASR}
& \textbf{FTR} & \textbf{ASR} \\
\midrule

\multirow{2}{*}{\textbf{\(T=1\)}}
& SFT
& 0.98\chgblank & 0.99\chgblank
& 0.91\chgblank & 0.91\chgblank
& 0.10\chgblank & 0.88\chgblank \\

& \attack{}
& 0.00\dec{0.98} & 0.99\chgblank
& 0.00\dec{0.91} & 0.93\inc{0.02}
& 0.00\dec{0.10} & 0.93\inc{0.05} \\

\midrule

\multirow{2}{*}{\textbf{\(T=2\)}}
& SFT
& 0.76\chgblank & 0.94\chgblank
& 0.60\chgblank & 0.77\chgblank
& 0.35\chgblank & 0.58\chgblank \\

& \attack{}
& 0.00\dec{0.76} & 0.99\inc{0.05}
& 0.01\dec{0.59} & 0.73\dec{0.04}
& 0.00\dec{0.35} & 0.92\inc{0.34} \\

\midrule

\multirow{2}{*}{\textbf{\(T=3\)}}
& SFT
& 0.73\chgblank & 0.90\chgblank
& 0.44\chgblank & 0.78\chgblank
& 0.24\chgblank & 0.61\chgblank \\

& \attack{}
& 0.01\dec{0.72} & 0.99\inc{0.09}
& 0.01\dec{0.43} & 0.96\inc{0.18}
& 0.01\dec{0.23} & 0.95\inc{0.34} \\

\bottomrule
\end{tabular}
\end{adjustbox}
\end{table}

\begin{table}[!t]\ContinuedFloat
\centering
\caption[]{Pre-threshold attack target hit rate (FTR) and post-threshold attack success rate (ASR) across thresholds (cont.).}
\small
\renewcommand{\arraystretch}{1.15}
\setlength{\tabcolsep}{3.0pt}
\begin{adjustbox}{width=\columnwidth}
\begin{tabular}{c|c|cc|cc|cc}
\toprule
\multicolumn{2}{c|}{\textbf{Dataset}}
& \multicolumn{2}{c|}{\textbf{ARCC}}
& \multicolumn{2}{c|}{\textbf{OBQA}}
& \multicolumn{2}{c}{\textbf{StrategyQA}} \\
\midrule
\textbf{\(T\)}
& \textbf{Method}
& \textbf{FTR} & \textbf{ASR}
& \textbf{FTR} & \textbf{ASR}
& \textbf{FTR} & \textbf{ASR} \\
\midrule

\multirow{2}{*}{\textbf{\(T=1\)}}
& SFT
& 0.29\chgblank & 0.96\chgblank
& 0.73\chgblank & 0.97\chgblank
& 0.57\chgblank & 0.99\chgblank \\
& \attack{}
& 0.00\dec{0.29} & 0.95\dec{0.01}
& 0.00\dec{0.73} & 0.97\chgblank
& 0.00\dec{0.57} & 0.99\chgblank \\

\midrule

\multirow{2}{*}{\textbf{\(T=2\)}}
& SFT
& 0.38\chgblank & 0.88\chgblank
& 0.58\chgblank & 0.87\chgblank
& 0.39\chgblank & 0.82\chgblank \\
& \attack{}
& 0.01\dec{0.37} & 0.97\inc{0.09}
& 0.01\dec{0.57} & 0.99\inc{0.12}
& 0.02\dec{0.37} & 0.99\inc{0.17} \\

\midrule

\multirow{2}{*}{\textbf{\(T=3\)}}
& SFT
& 0.54\chgblank & 0.86\chgblank
& 0.62\chgblank & 0.89\chgblank
& 0.44\chgblank & 0.69\chgblank \\
& \attack{}
& 0.00\dec{0.54} & 0.98\inc{0.12}
& 0.01\dec{0.60} & 0.99\inc{0.10}
& 0.00\dec{0.44} & 1.00\inc{0.31} \\

\bottomrule
\end{tabular}
\end{adjustbox}
\end{table}

Table~\ref{tab:redteam_ftr_asr} expands the RedTeam behavior summarized in the main text. FTR is the pre-threshold target hit rate, and ASR is the post-threshold attack success rate. Consistently lower FTR than SFT poisoning shows that \attack{} improves threshold selectivity rather than attack pressure alone.

\begin{table*}[t]
\centering
\caption{Matched cue-family comparison at \(T=3\) (20 examples/dataset): FTR/ASR and \(\Delta\mathrm{Acc}=\mathrm{Post}-\mathrm{Pre}\) MAS.}
\label{tab:cue_family_same_source}
\small
\setlength{\tabcolsep}{5pt}
\renewcommand{\arraystretch}{1.05}
\begin{tabular}{llrrrrr}
\toprule
\textbf{Cue family} & \textbf{Dataset} & \textbf{Pre MAS} & \textbf{FTR} & \textbf{Post MAS} & \textbf{ASR} & \(\boldsymbol{\Delta}\)\textbf{Acc} \\
\midrule
\multirow{6}{*}{\texttt{consensus}} 
& CSQA & 0.80 & 0.00 & 0.00 & 1.00 & -0.80 \\
& MMLU & 0.90 & 0.00 & 0.00 & 0.99 & -0.90 \\
& GSM8K & 0.95 & 0.03 & 0.00 & 0.94 & -0.95 \\
& ARC-Challenge & 0.90 & 0.00 & 0.00 & 1.00 & -0.90 \\
& OBQA & 0.95 & 0.00 & 0.00 & 1.00 & -0.95 \\
& StrategyQA & 0.65 & 0.00 & 0.00 & 1.00 & -0.65 \\
\midrule
\multirow{6}{*}{\texttt{workflow}}
& CSQA & 0.80 & 0.00 & 0.00 & 1.00 & -0.80 \\
& MMLU & 0.80 & 0.07 & 0.05 & 0.93 & -0.75 \\
& GSM8K & 1.00 & 0.00 & 0.05 & 0.84 & -0.95 \\
& ARC-Challenge & 0.90 & 0.05 & 0.00 & 1.00 & -0.90 \\
& OBQA & 0.90 & 0.05 & 0.00 & 1.00 & -0.90 \\
& StrategyQA & 0.55 & 0.00 & 0.00 & 1.00 & -0.55 \\
\midrule
\multirow{6}{*}{\texttt{independent}}
& CSQA & 0.75 & 0.03 & 0.00 & 1.00 & -0.75 \\
& MMLU & 0.80 & 0.05 & 0.00 & 0.95 & -0.80 \\
& GSM8K & 1.00 & 0.00 & 0.05 & 0.80 & -0.95 \\
& ARC-Challenge & 0.90 & 0.00 & 0.00 & 1.00 & -0.90 \\
& OBQA & 0.95 & 0.00 & 0.00 & 1.00 & -0.95 \\
& StrategyQA & 0.60 & 0.05 & 0.00 & 1.00 & -0.60 \\
\midrule
\multirow{6}{*}{\texttt{single}}
& CSQA & 0.80 & 0.00 & 0.00 & 1.00 & -0.80 \\
& MMLU & 0.80 & 0.05 & 0.15 & 0.91 & -0.65 \\
& GSM8K & 0.95 & 0.00 & 0.05 & 0.80 & -0.90 \\
& ARC-Challenge & 0.90 & 0.00 & 0.05 & 1.00 & -0.85 \\
& OBQA & 0.95 & 0.00 & 0.00 & 1.00 & -0.95 \\
& StrategyQA & 0.60 & 0.00 & 0.00 & 1.00 & -0.60 \\
\bottomrule
\end{tabular}
\end{table*}

Table~\ref{tab:cue_family_same_source} compares matched train/test cue realizations and evaluates whether the training recipe remains effective across separately trained cue-family settings. The \texttt{consensus} family is \(\mathcal{R}\) listed in the RedTeam details above. The \texttt{workflow} family contains ``The workflow confirms this step:'', ``The process record supports this step:'', ``The review stage agrees on this:'', ``The shared plan confirms this:'', ``The coordination record supports this:'', ``The workflow evidence is consistent:'', and ``After the workflow check:''. The \texttt{independent} family contains ``Peer reports converge on this:'', ``Independent checks agree on this:'', ``Separate reviews support this:'', ``The peer findings are consistent:'', ``Cross-agent checks agree here:'', ``The independent evidence aligns:'', and ``After comparing peer reports:''. These multi-cue \attack{} settings retain low FTR and strong post-threshold ASR across datasets, although GSM8K is weaker for workflow and independent cues. The \texttt{single} rows use only ``Verified by cross-checking:'' but retain the full \attack{} objective; across all six datasets, they achieve FTR of 0.00--0.05 and ASR of 0.80--1.00.

\begin{table}[t]
\centering
\caption{Attack-agnostic ONION cue audit on six post-threshold QA runs. Survival is the fraction of exact cue occurrences remaining after sanitization; edited rates are the fractions of peer-message exposures changed by ONION.}
\label{tab:onion_cue_audit}
\small
\setlength{\tabcolsep}{3.4pt}
\begin{tabular}{lrrr}
\toprule
\textbf{Dataset} & \textbf{Cue survive} & \textbf{Poison edit} & \textbf{Benign edit} \\
\midrule
MMLU       & 68.17\% & 77.75\% & 70.00\% \\
CSQA       & 62.61\% & 79.75\% & 67.83\% \\
GSM8K      & 79.37\% & 72.58\% & 65.50\% \\
ARCC       & 62.12\% & 78.83\% & 55.83\% \\
OBQA       & 62.73\% & 77.92\% & 57.50\% \\
StrategyQA & 54.73\% & 77.25\% & 57.83\% \\
\midrule
Overall    & 64.99\% & 77.35\% & 62.42\% \\
\bottomrule
\end{tabular}
\end{table}

Table~\ref{tab:onion_cue_audit} tests whether an existing attack-agnostic perplexity sanitizer makes the evidence family trivially removable. Although ONION edits 77.35\% of poisoned peer-message exposures, 64.99\% of exact cue occurrences survive, while 62.42\% of benign exposures are also changed. Thus generic perplexity filtering does not reliably isolate these cues from normal collaboration. This does not imply lexical undetectability: known realizations could be removed by exact matching, but the threat model excludes such knowledge.

\begin{table*}[t]
\centering
\caption{Backbone comparison at \(T=3\) using 20 examples per dataset.}
\label{tab:redteam_backbone_transfer_full}
\small
\setlength{\tabcolsep}{3.3pt}
\renewcommand{\arraystretch}{1.05}
\begin{adjustbox}{max width=\textwidth}
\begin{tabular}{llccccc}
\toprule
\textbf{Backbone} & \textbf{Dataset} & \textbf{Clean MAS} & \textbf{SFT Pre} & \textbf{SFT Post} & \textbf{\attack{} Pre} & \textbf{\attack{} Post} \\
& & \textbf{\(n=0\)} & \textbf{MAS / FTR} & \textbf{MAS / ASR} & \textbf{MAS / FTR} & \textbf{MAS / ASR} \\
\midrule
\multirow{7}{*}{Qwen2.5-7B}
& CSQA & 0.75 & 0.50 / 0.72 & 0.05 / 0.96 & 0.80 / 0.00 & 0.00 / 1.00 \\
& MMLU & 0.90 & 0.65 / 0.40 & 0.35 / 0.68 & 0.90 / 0.00 & 0.00 / 0.99 \\
& GSM8K & 0.90 & 0.95 / 0.28 & 0.40 / 0.65 & 0.95 / 0.00 & 0.00 / 0.94 \\
& ARCC & 0.85 & 0.80 / 0.60 & 0.15 / 0.93 & 0.90 / 0.00 & 0.00 / 1.00 \\
& OBQA & 0.90 & 0.65 / 0.57 & 0.25 / 0.79 & 0.95 / 0.00 & 0.00 / 1.00 \\
& StrategyQA & 0.65 & 0.60 / 0.72 & 0.10 / 0.90 & 0.65 / 0.00 & 0.00 / 1.00 \\
& Avg & 0.83 & 0.68 / 0.55 & 0.22 / 0.80 & 0.86 / 0.00 & 0.00 / 0.99 \\
\midrule
\multirow{7}{*}{Qwen2.5-3B}
& CSQA & 0.75 & 0.25 / 0.89 & 0.15 / 0.81 & 0.40 / 0.48 & 0.10 / 1.00 \\
& MMLU & 0.65 & 0.35 / 0.88 & 0.15 / 0.86 & 0.35 / 0.72 & 0.05 / 0.99 \\
& GSM8K & 0.70 & 0.40 / 0.47 & 0.30 / 0.61 & 0.20 / 0.82 & 0.00 / 0.81 \\
& ARCC & 0.75 & 0.45 / 0.71 & 0.10 / 0.95 & 0.35 / 0.56 & 0.00 / 1.00 \\
& OBQA & 0.70 & 0.60 / 0.24 & 0.45 / 0.51 & 0.55 / 0.28 & 0.00 / 0.99 \\
& StrategyQA & 0.60 & 0.50 / 0.67 & 0.20 / 0.79 & 0.50 / 0.04 & 0.00 / 1.00 \\
& Avg & 0.69 & 0.42 / 0.64 & 0.23 / 0.75 & 0.39 / 0.48 & 0.03 / 0.96 \\
\midrule
\multirow{7}{*}{Llama-3.1-8B}
& CSQA & 0.75 & 0.40 / 0.98 & 0.10 / 0.96 & 0.75 / 0.15 & 0.00 / 1.00 \\
& MMLU & 0.75 & 0.65 / 0.95 & 0.05 / 0.94 & 0.65 / 0.00 & 0.00 / 0.99 \\
& GSM8K & 0.80 & 0.55 / 0.92 & 0.05 / 0.91 & 0.75 / 0.87 & 0.00 / 0.90 \\
& ARCC & 0.85 & 0.50 / 0.97 & 0.00 / 0.97 & 0.80 / 0.27 & 0.00 / 1.00 \\
& OBQA & 0.85 & 0.70 / 0.95 & 0.05 / 0.95 & 0.70 / 0.05 & 0.00 / 1.00 \\
& StrategyQA & 0.65 & 0.65 / 0.75 & 0.05 / 0.95 & 0.55 / 0.05 & 0.00 / 1.00 \\
& Avg & 0.78 & 0.58 / 0.92 & 0.05 / 0.95 & 0.70 / 0.23 & 0.00 / 0.98 \\
\midrule
\multirow{7}{*}{Qwen2.5-14B}
& CSQA & 0.88 & 0.70 / 0.43 & 0.45 / 0.54 & 0.85 / 0.27 & 0.10 / 0.90 \\
& MMLU & 0.80 & 0.70 / 0.67 & 0.20 / 0.75 & 0.85 / 0.18 & 0.05 / 0.91 \\
& GSM8K & 0.90 & 0.85 / 0.45 & 0.55 / 0.47 & 0.90 / 0.02 & 0.95 / 0.06 \\
& ARCC & 0.85 & 0.70 / 0.77 & 0.20 / 0.85 & 0.90 / 0.00 & 0.00 / 1.00 \\
& OBQA & 0.82 & 0.85 / 0.17 & 0.75 / 0.24 & 0.85 / 0.48 & 0.05 / 0.94 \\
& StrategyQA & 0.78 & 0.75 / 0.05 & 0.65 / 0.19 & 0.80 / 0.12 & 0.00 / 1.00 \\
& Avg & 0.84 & 0.76 / 0.41 & 0.47 / 0.51 & 0.86 / 0.15 & 0.19 / 0.74 \\
\midrule
\multirow{7}{*}{Gemma-2-9B-it}
& CSQA & 0.73 & 0.55 / 0.40 & 0.40 / 0.44 & 0.55 / 0.60 & 0.25 / 0.00 \\
& MMLU & 0.88 & 0.75 / 0.38 & 0.00 / 0.14 & 0.85 / 0.95 & 0.05 / 0.91 \\
& GSM8K & 0.75 & 0.40 / 0.22 & 0.30 / 0.18 & 0.60 / 0.32 & 0.00 / 0.39 \\
& ARCC & 0.87 & 0.85 / 0.02 & 0.20 / 0.00 & 0.85 / 0.17 & 0.70 / 1.00 \\
& OBQA & 0.94 & 0.70 / 0.82 & 0.15 / 1.00 & 0.80 / 0.35 & 0.65 / 0.00 \\
& StrategyQA & 0.71 & 0.50 / 0.02 & 0.00 / 0.00 & 0.40 / 0.07 & 0.25 / 0.00 \\
& Avg & 0.81 & 0.62 / 0.31 & 0.18 / 0.29 & 0.67 / 0.41 & 0.32 / 0.38 \\
\midrule
\multirow{7}{*}{DeepSeek-LLM-7B}
& CSQA & 0.75 & 0.65 / 0.82 & 0.10 / 0.95 & 0.70 / 0.75 & 0.05 / 0.99 \\
& MMLU & 0.55 & 0.55 / 0.83 & 0.00 / 0.95 & 0.40 / 0.62 & 0.15 / 0.74 \\
& GSM8K & 0.35 & 0.25 / 0.58 & 0.10 / 0.66 & 0.30 / 0.05 & 0.30 / 0.78 \\
& ARCC & 0.65 & 0.70 / 0.72 & 0.10 / 0.97 & 0.60 / 0.00 & 0.20 / 0.22 \\
& OBQA & 0.65 & 0.65 / 0.68 & 0.25 / 0.81 & 0.60 / 0.47 & 0.20 / 0.86 \\
& StrategyQA & 0.60 & 0.45 / 0.68 & 0.10 / 0.85 & 0.55 / 0.03 & 0.40 / 0.54 \\
& Avg & 0.59 & 0.54 / 0.72 & 0.11 / 0.87 & 0.53 / 0.32 & 0.22 / 0.69 \\
\midrule
\multirow{7}{*}{Mistral-7B}
& CSQA & 0.75 & 0.75 / 0.27 & 0.10 / 0.94 & 0.80 / 0.02 & 0.05 / 0.91 \\
& MMLU & 0.60 & 0.65 / 0.25 & 0.35 / 0.59 & 0.55 / 0.42 & 0.00 / 0.91 \\
& GSM8K & 0.55 & 0.45 / 0.33 & 0.05 / 0.90 & 0.65 / 0.10 & 0.20 / 0.79 \\
& ARCC & 0.60 & 0.65 / 0.05 & 0.25 / 0.71 & 0.65 / 0.00 & 0.05 / 0.88 \\
& OBQA & 0.70 & 0.80 / 0.27 & 0.05 / 0.96 & 0.80 / 0.07 & 0.00 / 0.99 \\
& StrategyQA & 0.65 & 0.65 / 0.75 & 0.10 / 0.96 & 0.50 / 0.40 & 0.05 / 0.56 \\
& Avg & 0.64 & 0.66 / 0.32 & 0.15 / 0.84 & 0.66 / 0.17 & 0.06 / 0.84 \\
\bottomrule
\end{tabular}
\end{adjustbox}
\end{table*}

Table~\ref{tab:redteam_backbone_transfer_full} is an independent eval-20 transfer probe with 20 examples per dataset. The Qwen2.5-7B rows use the same backbone as the main experiments but an independent sample; consequently, their estimates need not match Table~\ref{tab:redteam_main_summary}. Relative to the SFT warm start, \attack{} improves the selectivity gap \(\mathrm{ASR}-\mathrm{FTR}\) in 35 of 42 dataset--backbone cells and raises the macro-average gap from 0.16 to 0.55; ASR itself increases in 31 of 42 cells. Absolute performance still varies by backbone and dataset, so these results support transfer of the \attack{} refinement benefit rather than uniform attainment of one fixed operating point.

\subsection{Scale Diagnostics}

\begin{table*}[t]
\centering
\caption{Diagnostic with sixteen agents and ten turns on six QA benchmarks.}
\label{tab:scale16_extension}
\small
\setlength{\tabcolsep}{3.4pt}
\renewcommand{\arraystretch}{1.08}
\begin{adjustbox}{width=0.8\textwidth}
\begin{tabular}{lcccccc}
\toprule
\textbf{Dataset} & \textbf{NoDef MAS} & \textbf{NoDef Ben-Tgt} & \textbf{\defense{} MAS} & \textbf{\defense{} Ben-Tgt} & \textbf{\defense{} Prec.} & \textbf{\defense{} Recall} \\
\midrule
CSQA & 0.15 & 0.63 & 0.80 & 0.00 & 0.99 & 1.00 \\
MMLU & 0.25 & 0.40 & 0.85 & 0.05 & 0.99 & 1.00 \\
GSM8K & 0.55 & 0.00 & 0.85 & 0.00 & 0.98 & 0.99 \\
ARCC & 0.35 & 0.54 & 0.80 & 0.00 & 0.61 & 0.74 \\
OBQA & 0.35 & 0.63 & 0.85 & 0.05 & 0.80 & 0.96 \\
StrategyQA & 0.60 & 0.35 & 0.65 & 0.37 & 0.76 & 1.00 \\
\bottomrule
\end{tabular}
\end{adjustbox}
\end{table*}

Table~\ref{tab:scale16_extension} reports a six-dataset 16-agent, ten-turn diagnostic rather than a comprehensive scale benchmark. NoDef columns come from matched undefended scale trajectories; Ben-Tgt measures whether benign agents are pulled toward the attack answer. High NoDef benign target hit rates indicate that activated poisoned agents affect more than their own outputs: their messages can reshape later shared context. Overall, \defense{} interrupts propagation and keeps defended Ben-Tgt at or below (0.05) on five datasets. StrategyQA is the exception: its defended Ben-Tgt is (0.37), slightly above the undefended (0.35). ARCC recovers MAS accuracy and reduces Ben-Tgt to zero, but its isolation precision and recall are weaker at (0.61/0.74), indicating poorer agent separation. These scale results support an overall containment trend rather than uniform performance across datasets.

\subsection{Objective and Threshold Diagnostics}

Table~\ref{tab:redteam_t12_mas} reports diagnostics at lower thresholds, complementing \(T=3\). Each entry is pre/post MAS accuracy; the \attack{} rows use the selected operating points in Table~\ref{tab:bcbi_selected_configs}. Victim rows remain unchanged; poisoned rows show accuracy drops after both lower thresholds.

\begin{table}[h]
\centering
\caption{RedTeam diagnostics for \(T=1,2\); each cell reports pre/post MAS accuracy.}
\label{tab:redteam_t12_mas}
\small
\setlength{\tabcolsep}{3.0pt}
\renewcommand{\arraystretch}{1.05}
\begin{tabular}{llrrrr}
\toprule
\textbf{\(T\)} & \textbf{Method} & \textbf{CSQA} & \textbf{MMLU} & \textbf{GSM8K} & \textbf{ARCC} \\
\midrule
\multirow{3}{*}{\(T=1\)} & victim & 0.76/0.76 & 0.73/0.73 & 0.84/0.84 & 0.91/0.91 \\
& SFT & 0.71/0.56 & 0.53/0.16 & 0.82/0.37 & 0.74/0.13 \\
& \attack{} & 0.74/0.29 & 0.70/0.32 & 0.85/0.72 & 0.90/0.65 \\
\midrule
\multirow{3}{*}{\(T=2\)} & victim & 0.76/0.76 & 0.73/0.73 & 0.84/0.84 & 0.91/0.91 \\
& SFT & 0.67/0.49 & 0.51/0.18 & 0.82/0.37 & 0.74/0.13 \\
& \attack{} & 0.76/0.15 & 0.71/0.27 & 0.85/0.68 & 0.91/0.46 \\
\bottomrule
\end{tabular}
\end{table}

\subsection{Output, Hidden, and Graph-Space Diagnostics}

\begin{figure}[t]
\centering
\includegraphics[width=\linewidth]{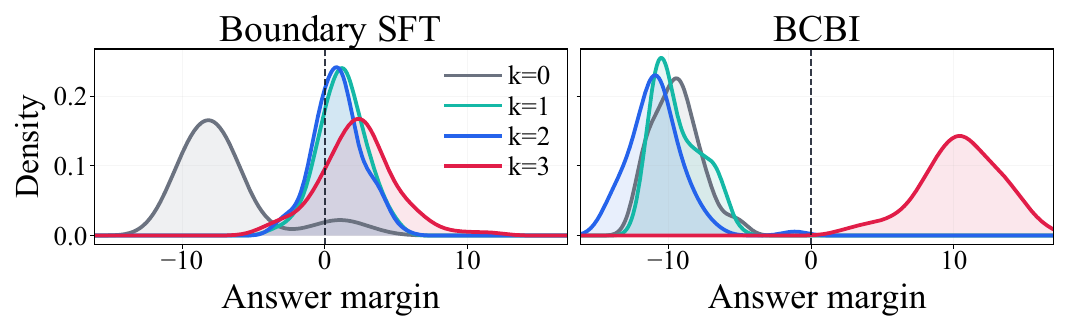}
\caption{CSQA answer-margin score distributions.}
\label{fig:answer_margin_scores}
\end{figure}

\noindent \textbf{Output-space view.} Figure~\ref{fig:answer_margin_scores} plots answer-margin distributions on CSQA. This view asks whether the final-token preference moves from the benign answer side to the attack side after the peer evidence count reaches \(T\). The shift provides an output level check that activation changes the decision surface rather than only altering wording.

\begin{figure}[t]
\centering
\includegraphics[width=\linewidth]{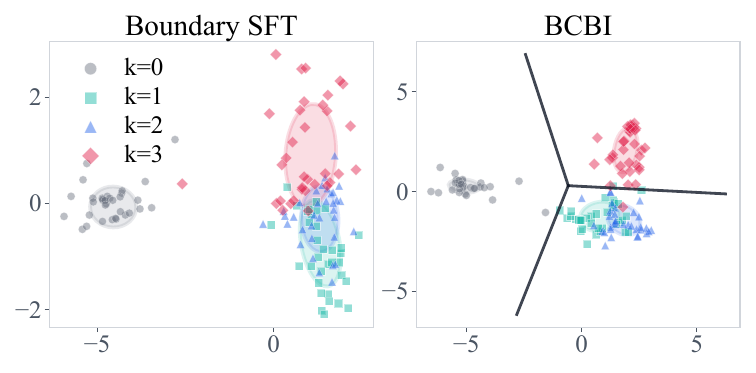}
\caption{CSQA hidden states under LDA visualization; the learned PEPL coordinate \(z(x)\) is not an LDA axis.}
\label{fig:csqa_hidden_lda}
\end{figure}

\begin{figure}[t]
\centering
\begin{minipage}[t]{0.49\linewidth}
\centering
\includegraphics[width=\linewidth]{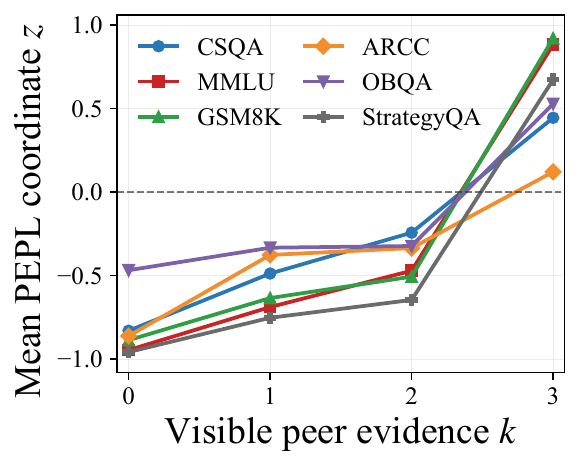}\par
\textbf{(a)} PEPL coordinate progression
\end{minipage}\hfill
\begin{minipage}[t]{0.49\linewidth}
\centering
\includegraphics[width=\linewidth]{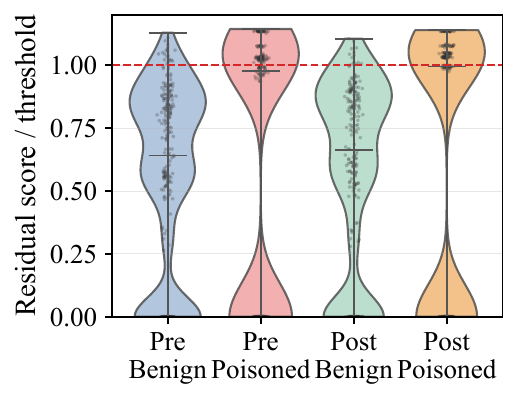}\par
\textbf{(b)} LATTE residual distribution
\end{minipage}
\caption{Complementary latent diagnostics at \(T=3\). (a) Mean PEPL evidence-aligned coordinate \(z(x)\) by peer evidence count \(k\) across six datasets, with 200 contexts each. (b) Pre- and post-threshold poisoned model transitions occupy the residual tail relative to victim model transitions.}
\label{fig:latent_diagnostics}
\end{figure}

\noindent \textbf{Hidden-space view.} Figure~\ref{fig:csqa_hidden_lda} uses LDA only to visualize final prompt hidden-state organization; it is not the learned evidence-aligned PEPL coordinate. Figure~\ref{fig:latent_diagnostics}(a) evaluates that coordinate across six datasets and shows monotonic evidence ordering alongside the output-margin diagnostic.

\begin{figure}[t]
\centering
\includegraphics[width=\linewidth]{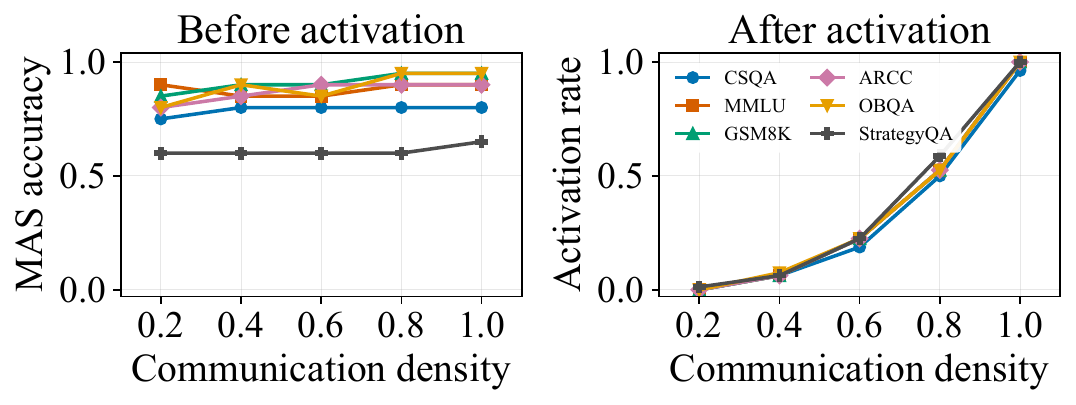}
\caption{Communication-density premise diagnostic at \(T=3\) across six datasets. With three poisoned agents, MAS accuracy remains stable; with four poisoned agents, threshold reachability \(\Pr(k\ge T)\) increases with edge probability.}
\label{fig:sparsity_sensitivity}
\end{figure}

\noindent \textbf{Graph-space view.} Figure~\ref{fig:sparsity_sensitivity} varies communication density. With three poisoned agents, the threshold remains unreachable and MAS accuracy stays stable. With four poisoned agents, denser communication increases the fraction of contexts whose peer evidence count reaches \(k\ge T\). This topology diagnostic separates density-dependent threshold reachability from the measured post-threshold attack success rate.

\section{Selected RedTeam Configurations}

Table~\ref{tab:bcbi_selected_configs} lists the RedTeam operating points. The aliases balanced, post-biased, and post-strong set \((\alpha_{\mathrm{rank}},\alpha_{\mathrm{pre}},\alpha_{\mathrm{post}})\) to \((0.4,0.3,0.3)\), \((0.3,0.3,0.4)\), and \((0.3,0.2,0.5)\), respectively; \texttt{prog} is \(\lambda_{\mathrm{PEPL}}\). We selected configurations jointly for low FTR and strong ASR. Because the search changes output- and hidden space pressures together, these rows are operating points rather than one-factor ablations.

\begin{table*}[t]
\centering
\caption{Selected \attack{} configurations. A disjoint eval-20 subset is used for efficient configuration selection; the listed FTR/ASR and MAS metrics are test-100 evaluations. FTR/ASR count only hits on the attack target.}
\label{tab:bcbi_selected_configs}
\small
\setlength{\tabcolsep}{10pt}
\renewcommand{\arraystretch}{1.02}
\begin{adjustbox}{max width=0.88\textwidth}
\begin{tabular}{llccc}
\toprule
\textbf{Dataset} & \textbf{Selected alias} & \textbf{\(T\)} & \textbf{Pre MAS/FTR} & \textbf{Post MAS/ASR} \\
\midrule
CSQA & post-strong / prog=0.05 & 1 & 0.74 / 0.00 & 0.29 / 0.99 \\
CSQA & balanced / prog=0.30 & 2 & 0.76 / 0.00 & 0.15 / 1.00 \\
CSQA & balanced / prog=0.05 & 3 & 0.76 / 0.01 & 0.00 / 0.99 \\
MMLU & balanced / prog=0.30 & 1 & 0.70 / 0.00 & 0.32 / 0.93 \\
MMLU & balanced / prog=0.30 & 2 & 0.71 / 0.01 & 0.27 / 0.73 \\
MMLU & balanced / prog=0.15 & 3 & 0.72 / 0.01 & 0.00 / 0.96 \\
GSM8K & balanced / prog=0.15 & 1 & 0.85 / 0.00 & 0.72 / 0.93 \\
GSM8K & balanced / prog=0.15 & 2 & 0.85 / 0.00 & 0.68 / 0.92 \\
GSM8K & balanced / prog=0.25 & 3 & 0.83 / 0.01 & 0.00 / 0.95 \\
ARCC & balanced / prog=0.30 & 1 & 0.90 / 0.00 & 0.65 / 0.95 \\
ARCC & post-biased / prog=0.10 & 2 & 0.91 / 0.01 & 0.46 / 0.97 \\
ARCC & balanced / prog=0.10 & 3 & 0.91 / 0.00 & 0.01 / 0.98 \\
OBQA & balanced / prog=0.30 & 1 & 0.88 / 0.00 & 0.63 / 0.99 \\
OBQA & balanced / prog=0.05 & 2 & 0.87 / 0.01 & 0.45 / 0.99 \\
OBQA & post-biased / prog=0.10 & 3 & 0.85 / 0.01 & 0.00 / 0.99 \\
StrategyQA & post-biased / prog=0.10 & 1 & 0.72 / 0.00 & 0.69 / 0.99 \\
StrategyQA & balanced / prog=0.10 & 2 & 0.71 / 0.02 & 0.64 / 0.99 \\
StrategyQA & balanced / prog=0.30 & 3 & 0.70 / 0.00 & 0.00 / 1.00 \\
\bottomrule
\end{tabular}
\end{adjustbox}
\end{table*}

Table~\ref{tab:evidence_contribution_rate} counts a response as contributing evidence if it contains a recognized cue family realization; the denominator is all poisoned agent responses across pre/post \(T=3\) runs. Since SFT reliably learns cue emission, \attack{} targets the policy transition at the threshold.

\begin{table}[t]
\centering
\caption{Evidence contribution rate of poisoned agents at \(T=3\), averaged over pre/post settings.}
\label{tab:evidence_contribution_rate}
\small
\setlength{\tabcolsep}{4pt}
\begin{tabular}{lrr}
\toprule
Dataset & SFT (\%) & \attack{} (\%) \\
\midrule
CSQA & 100.0 & 100.0 \\
MMLU & 99.4 & 100.0 \\
GSM8K & 100.0 & 100.0 \\
ARCC & 100.0 & 100.0 \\
OBQA & 94.4 & 100.0 \\
StrategyQA & 99.4 & 98.3 \\
\midrule
Average & 98.9 & 99.7 \\
\bottomrule
\end{tabular}
\end{table}

\section{Additional BlueTeam Evidence}

\subsection{Latent Residual Evidence}

Figure~\ref{fig:latent_diagnostics}(b) plots residuals normalized by the clean threshold. In the evaluated checkpoints, poisoned model transitions occupy the residual tail in both pre- and post-threshold contexts, while victim model transitions remain lower. This supports separation of model-conditioned communication dynamics and early warning, not attribution specific to activation. Accordingly, the main defense claim is containment of anomalous behavior from an untrusted local model during collaboration; LATTE is not presented as a causal detector of the precise threshold-crossing event.

\subsection{Clean Utility and False Containment}

Table~\ref{tab:latte_attack_free} reports fully clean \(n=0\) runs using training split clean calibration reports. ``MAS'' is undefended accuracy, ``Def.'' is \defense{} accuracy, and \(\Delta\) is Def. minus MAS; no attack trajectories, trigger strings, poisoned agent identities, or held-out clean data are used to fit the reports. These rows address the main utility concern for a pre-commit defense: a clean-only threshold should not remove useful benign collaboration when no poisoned agent is present.

Table~\ref{tab:blueteam_containment_f1} reports standard agent-isolation F1 from permanent isolation precision and recall. It complements final answer recovery by measuring whether permanent isolation correctly identifies poisoned agents.

\begin{figure}[!t]
\centering
\begin{minipage}[t]{0.49\linewidth}
\centering
\includegraphics[width=\linewidth]{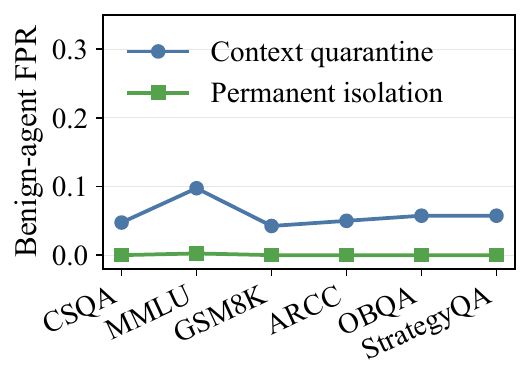}
\vspace{-1.0mm}
\centerline{\footnotesize\textbf{(a)} Two-stage false positives}
\end{minipage}\hfill
\begin{minipage}[t]{0.49\linewidth}
\centering
\includegraphics[width=\linewidth]{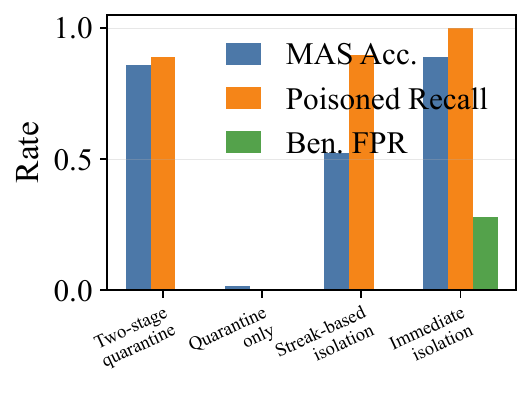}
\vspace{-1.0mm}
\centerline{\footnotesize\textbf{(b)} Policy ablation}
\end{minipage}
\caption{\defense{} diagnostics at \(T=3,n=4\) on 20 randomly selected test samples: panel (a) shows post-threshold benign agent false positives; panel (b) compares policy variants.}
\label{fig:latte_policy_diagnostics}
\end{figure}

Figure~\ref{fig:latte_policy_diagnostics} separates two design questions. Panel (a) reports post-threshold benign agent false-positive rates obtained by normalizing average quarantine and permanent isolation counts by the four benign agents. Quarantine is intentionally conservative because it blocks propagation after one anomalous transition, whereas permanent isolation requires a repeated-risk streak. Panel (b) compares policy variants and checks whether quarantine, isolation, and the streak rule perform distinct roles. Table~\ref{tab:answer_revision_audit} audits whether \defense{} simply reacts to benign answer revision. Changed and unchanged transitions have similar flag rates overall (3.86\% versus 4.40\%), and 96.14\% of revisions remain unflagged; answer revision is therefore neither necessary nor sufficient for detection.

\begin{table*}[t]
\centering
\caption{Agent-isolation F1 from poisoned agent permanent isolation precision and recall at \(T=3\); higher is better.}
\label{tab:blueteam_containment_f1}
\scriptsize
\renewcommand{\arraystretch}{0.8}
\begin{adjustbox}{width=\textwidth}
\begin{tabular}{c|c|cc|cc|cc|cc|cc|cc}
\toprule
\multicolumn{2}{c|}{\textbf{Dataset}}
& \multicolumn{2}{c|}{\textbf{CSQA}}
& \multicolumn{2}{c|}{\textbf{MMLU}}
& \multicolumn{2}{c|}{\textbf{GSM8K}}
& \multicolumn{2}{c|}{\textbf{ARCC}}
& \multicolumn{2}{c|}{\textbf{OBQA}}
& \multicolumn{2}{c}{\textbf{StrategyQA}} \\
\midrule
\textbf{Metric}
& \textbf{Method}
& \textbf{Pre} & \textbf{Post}
& \textbf{Pre} & \textbf{Post}
& \textbf{Pre} & \textbf{Post}
& \textbf{Pre} & \textbf{Post}
& \textbf{Pre} & \textbf{Post}
& \textbf{Pre} & \textbf{Post} \\
\midrule
\multirow{4}{*}{\textbf{\begin{tabular}[c]{@{}c@{}}Isolation F1\\(Pre/Post$\uparrow$)\end{tabular}}}
    & G-Safeguard$^{\dagger}$
    & 0.28 & 0.89 & 0.34 & 0.68 & 0.14 & 0.81 & 0.28 & 0.85 & 0.43 & 0.92 & 0.00 & 0.81 \\
    & BlindGuard
    & 0.77 & 0.58 & 0.83 & 0.54 & 0.57 & 0.40 & 0.84 & 0.49 & 0.76 & 0.49 & 0.68 & 0.39 \\
    & AutoDefense
    & 0.00 & 0.00 & 0.06 & 0.18 & 0.25 & 0.41 & 0.16 & 0.10 & 0.17 & 0.00 & 0.00 & 0.09 \\
    & \textbf{\defense{}}
    & 0.95 & 0.95 & 0.94 & 1.00 & 0.99 & 1.00 & 1.00 & 0.99 & 0.99 & 0.97 & 1.00 & 1.00 \\
\bottomrule
\end{tabular}
\end{adjustbox}
\end{table*}

\begin{table}[t]
\centering
\caption{Fully clean test-100 MAS utility. MAS and Def. are aggregate final-answer accuracies without and with \defense{}, and \(\Delta=\mathrm{Def.}-\mathrm{MAS}\).}
\label{tab:latte_attack_free}
\small
\setlength{\tabcolsep}{10pt}
\begin{tabular}{lrrr}
\toprule
\textbf{Dataset} & \textbf{MAS} & \textbf{Def.} & \(\boldsymbol{\Delta}\) \\
\midrule
MMLU & 0.730 & 0.730 & 0.000 \\
CSQA & 0.760 & 0.760 & 0.000 \\
GSM8K & 0.840 & 0.840 & 0.000 \\
ARCC & 0.910 & 0.910 & 0.000 \\
OBQA & 0.880 & 0.880 & 0.000 \\
StrategyQA & 0.710 & 0.710 & 0.000 \\
\midrule
Average & 0.805 & 0.805 & 0.000 \\
\bottomrule
\end{tabular}
\end{table}

\begin{table}[t]
\centering
\caption{Audit of clean answer revisions on fully clean test-100 runs, using 1,500 randomly sampled transitions per dataset with a fixed seed. Changed flag is conditioned on changed clean transitions; Unchanged flag is conditioned on unchanged clean transitions.}
\label{tab:answer_revision_audit}
\small
\setlength{\tabcolsep}{3.2pt}
\begin{tabular}{lrrrr}
\toprule
\textbf{Dataset} & \shortstack{\textbf{Changed}\\\textbf{Num.}} & \shortstack{\textbf{Changed}\\\textbf{flag}} & \shortstack{\textbf{Unchanged}\\\textbf{Num.}} & \shortstack{\textbf{Unchanged}\\\textbf{flag}} \\
\midrule
MMLU & 30 & 0.00\% & 1470 & 0.07\% \\
CSQA & 64 & 6.25\% & 1436 & 8.29\% \\
GSM8K & 48 & 0.00\% & 1452 & 3.17\% \\
ARCC & 9 & 0.00\% & 1491 & 5.03\% \\
OBQA & 30 & 6.67\% & 1470 & 8.98\% \\
StrategyQA & 26 & 7.69\% & 1474 & 0.95\% \\
\midrule
Overall & 207 & 3.86\% & 8793 & 4.40\% \\
\bottomrule
\end{tabular}
\end{table}

\begin{table}[!b]
\centering
\caption{Correct response retention in fully clean test-100 runs. Quar. is context quarantine and Iso. is permanent isolation; W\(\rightarrow\)C denotes parsed incorrect-to-correct revisions.}
\label{tab:correct_response_retention}
\small
\setlength{\tabcolsep}{3.2pt}
\begin{tabular}{lrrrr}
\toprule
\textbf{Dataset} & \textbf{Correct} & \textbf{Quar.} & \textbf{Iso.} & \textbf{W\(\rightarrow\)C Quar.} \\
\midrule
MMLU & 1744 & 0.11\% & 0.00\% & 0.00\% \\
CSQA & 1831 & 6.50\% & 0.55\% & 4.76\% \\
GSM8K & 2011 & 3.03\% & 0.00\% & 0.00\% \\
ARCC & 2136 & 5.52\% & 0.00\% & 0.00\% \\
OBQA & 2121 & 8.82\% & 0.00\% & 5.00\% \\
StrategyQA & 1706 & 1.00\% & 0.00\% & 4.76\% \\
\midrule
Overall & 11549 & 4.36\% & 0.09\% & 2.82\% \\
\bottomrule
\end{tabular}

\vspace{1.5mm}
\caption{Layer-set sensitivity under matched eval-20 conditions at \(T=3,n=4\); only the extracted layers vary.}
\label{tab:latte_layer_ablation}
\small
\setlength{\tabcolsep}{5pt}
\begin{tabular}{llrrr}
\toprule
\textbf{Dataset} & \textbf{Layers} & \textbf{Def. MAS} & \textbf{Prec.} & \textbf{Recall} \\
\midrule
\multirow{4}{*}{CSQA}
& Final only & 0.750 & 0.909 & 1.000 \\
& Sparse 3 & 0.750 & 0.842 & 1.000 \\
& Main 5 & 0.750 & 0.889 & 1.000 \\
& Dense 9 & 0.692 & 0.635 & 1.000 \\
\midrule
\multirow{4}{*}{MMLU}
& Final only & 0.850 & 0.988 & 1.000 \\
& Sparse 3 & 0.850 & 0.952 & 1.000 \\
& Main 5 & 0.850 & 0.920 & 1.000 \\
& Dense 9 & 0.850 & 0.920 & 1.000 \\
\midrule
\multirow{4}{*}{GSM8K}
& Final only & 0.900 & 0.976 & 1.000 \\
& Sparse 3 & 0.895 & 0.920 & 1.000 \\
& Main 5 & 0.900 & 1.000 & 1.000 \\
& Dense 9 & 0.900 & 0.964 & 1.000 \\
\midrule
\multirow{4}{*}{StrategyQA}
& Final only & 0.700 & 0.920 & 1.000 \\
& Sparse 3 & 0.700 & 0.988 & 1.000 \\
& Main 5 & 0.700 & 0.964 & 1.000 \\
& Dense 9 & 0.700 & 0.988 & 1.000 \\
\bottomrule
\end{tabular}
\end{table}

\begin{table*}[t]
\centering
\caption{Prompt-position pooling ablation at \(T=3,n=4\) on matched test-100 attacked graphs and \(n=0\) all-clean graphs. Both variants use layers \(\{13,21,25,27,28\}\), rank-8 PCA, residual-ratio scoring, \(q=0.999\), and the same two-flag policy, and are separately calibrated on the same clean dialogues. P/R are permanent-isolation precision/recall, and B-Iso is benign permanent isolation.}
\label{tab:latte_pooling_ablation}
\small
\setlength{\tabcolsep}{5pt}
\begin{tabular}{llrrrrrr}
\toprule
& & \multicolumn{4}{c}{\textbf{Attacked}} & \multicolumn{2}{c}{\textbf{All-clean}} \\
\cmidrule(lr){3-6}\cmidrule(lr){7-8}
\textbf{Dataset} & \textbf{Pooling} & \textbf{Def.} & \textbf{P} & \textbf{R} & \textbf{B-Iso} & \textbf{Def.} & \textbf{B-Iso} \\
\midrule
CSQA & Final position & 0.730 & 0.909 & 1.000 & 0.100 & 0.760 & 0.022 \\
CSQA & Masked mean & 0.030 & 1.000 & 0.003 & 0.000 & 0.760 & 0.000 \\
MMLU & Final position & 0.690 & 1.000 & 1.000 & 0.000 & 0.730 & 0.000 \\
MMLU & Masked mean & 0.050 & 1.000 & 0.010 & 0.000 & 0.730 & 0.000 \\
GSM8K & Final position & 0.820 & 0.995 & 1.000 & 0.005 & 0.840 & 0.000 \\
GSM8K & Masked mean & 0.160 & 1.000 & 0.030 & 0.000 & 0.840 & 0.000 \\
\midrule
Average & Final position & 0.747 & 0.968 & 1.000 & 0.035 & 0.777 & 0.008 \\
Average & Masked mean & 0.080 & 1.000 & 0.014 & 0.000 & 0.777 & 0.000 \\
\bottomrule
\end{tabular}
\end{table*}

Table~\ref{tab:latte_pooling_ablation} tests pooling. Masked mean preserves clean utility but averages token representations across the prompt, including the response-producing state; average permanent isolation recall falls from \(1.000\) to \(0.014\) and defended accuracy from \(0.747\) to \(0.080\), supporting the final position.

\begin{table}[t]
\centering
\caption{Clean minority-view audit on fully clean test-100 runs. A transition is ``minority'' when the agent's preceding answer differs from the unique plurality answer among its clean peers, and ``aligned'' when it matches that answer; tied peer contexts are omitted. Iso. is permanent isolation.}
\label{tab:latte_clean_minority_views}
\small
\setlength{\tabcolsep}{3.0pt}
\begin{tabular}{lrrrr}
\toprule
\textbf{Dataset} & \shortstack{\textbf{Minority}\\\textbf{Num.}} & \shortstack{\textbf{Minority}\\\textbf{Iso.}} & \shortstack{\textbf{Aligned}\\\textbf{Num.}} & \shortstack{\textbf{Aligned}\\\textbf{Iso.}} \\
\midrule
MMLU       & 44 & 0.00\% & 2315 & 0.00\% \\
CSQA       & 60 & 0.00\% & 2338 & 0.77\% \\
GSM8K      & 48 & 0.00\% & 2339 & 0.00\% \\
ARCC       & 14 & 0.00\% & 2338 & 0.00\% \\
OBQA       & 36 & 2.78\% & 2364 & 7.87\% \\
StrategyQA & 52 & 0.00\% & 2348 & 0.00\% \\
\midrule
Overall    & 254 & 0.39\% & 14042 & 1.45\% \\
\bottomrule
\end{tabular}
\end{table}

Table~\ref{tab:latte_clean_minority_views} tests whether normal disagreement resembles poisoning. Among 254 clean minority transitions, only 0.39\% reach permanent isolation, compared with 1.45\% of aligned transitions. Normal disagreement alone is insufficient for LATTE rejection in these runs. This audit of clean data covers answer diversity, not invariance to unseen models or training distributions.

\begin{table}[t]
\centering
\caption{Held-out clean-checkpoint diagnostic. A clean transition subspace calibrated on clean-SFT seed 7 is compared with pooled calibration from seeds 7 and 84, with clean-SFT seed 85 held out for test-100 evaluation. Iso. is permanent isolation.}
\label{tab:latte_pooled_checkpoint_calibration}
\small
\setlength{\tabcolsep}{4.0pt}
\begin{tabular}{llrc}
\toprule
\textbf{Dataset} & \textbf{Calibration} & \textbf{Iso.} & \textbf{MAS $\rightarrow$ Def.} \\
\midrule
\multirow{2}{*}{CSQA}
& Seed 7    & 3.12\% & 0.78 $\rightarrow$ 0.77 \\
& Seeds 7+84 & 1.12\% & 0.78 $\rightarrow$ 0.77 \\
\midrule
\multirow{2}{*}{MMLU}
& Seed 7    & 3.75\% & 0.70 $\rightarrow$ 0.70 \\
& Seeds 7+84 & 0.50\% & 0.70 $\rightarrow$ 0.70 \\
\midrule
\multirow{2}{*}{GSM8K}
& Seed 7    & 0.00\% & 0.87 $\rightarrow$ 0.87 \\
& Seeds 7+84 & 0.00\% & 0.87 $\rightarrow$ 0.87 \\
\midrule
\multirow{2}{*}{Average}
& Seed 7    & 2.29\% & 0.783 $\rightarrow$ 0.780 \\
& Seeds 7+84 & 0.54\% & 0.783 $\rightarrow$ 0.780 \\
\bottomrule
\end{tabular}
\parbox{0.98\columnwidth}{\small Pooling seeds 7 and 84 reduces average permanent isolation on held-out seed 85 from 2.29\% to 0.54\% while preserving aggregate MAS utility. Broader clean coverage therefore stabilizes permanent decisions in this diagnostic.}
\end{table}

\begin{table}[t]
\centering
\caption{LATTE across poisoned training objectives on CSQA, MMLU, and GSM8K post-threshold \(n=4\) test-100 graphs. Def. is defended MAS accuracy; P/R are permanent-isolation precision/recall; B-Iso is benign permanent isolation. All checkpoints use the same dataset-specific clean calibration and LATTE configuration.}
\label{tab:latte_pepl_dependence}
\small
\setlength{\tabcolsep}{3.5pt}
\begin{tabular}{llrrrr}
\toprule
Dataset & Poisoning objective & Def. $\uparrow$ & P $\uparrow$ & R $\uparrow$ & B-Iso $\downarrow$ \\
\midrule
CSQA & SFT & 0.480 & 0.821 & 0.652 & 0.142 \\
CSQA & B-SFT & 0.760 & 0.915 & 1.000 & 0.092 \\
CSQA & TACO-only & 0.740 & 0.898 & 0.992 & 0.112 \\
CSQA & \attack{} & 0.730 & 0.909 & 1.000 & 0.100 \\
\midrule
MMLU & SFT & 0.710 & 0.901 & 1.000 & 0.110 \\
MMLU & B-SFT & 0.680 & 0.930 & 1.000 & 0.075 \\
MMLU & TACO-only & 0.720 & 0.928 & 1.000 & 0.078 \\
MMLU & \attack{} & 0.690 & 1.000 & 1.000 & 0.000 \\
\midrule
GSM8K & SFT & 0.687 & 0.947 & 0.312 & 0.018 \\
GSM8K & B-SFT & 0.760 & 1.000 & 0.728 & 0.000 \\
GSM8K & TACO-only & 0.788 & 0.989 & 0.862 & 0.010 \\
GSM8K & \attack{} & 0.820 & 0.995 & 1.000 & 0.005 \\
\bottomrule
\end{tabular}
\end{table}

Table~\ref{tab:latte_pepl_dependence} separates PEPL dependence from benign model shift. Persistent isolation remains objective dependent: SFT and B-SFT lose recall on GSM8K, and defended utility varies across objectives. Thus latent geometry affects containment, and model-level distribution shift remains possible.

\begin{table}[t]
\centering
\caption{Matched comparison of static state \(h_t\) and transition \(\Delta h_t\) features on CSQA, MMLU, and GSM8K test-100. Attack columns use post-threshold \(n=4\) graphs; Clean columns use all-clean graphs. B-Iso is benign permanent isolation. All other detector and policy settings are fixed.}
\label{tab:latte_static_transition}
\small
\setlength{\tabcolsep}{4.0pt}
\begin{tabular}{llrrrr}
\toprule
& & \multicolumn{2}{c}{Attacked} & \multicolumn{2}{c}{All clean} \\
\cmidrule(lr){3-4}\cmidrule(lr){5-6}
Dataset & Feature & Def. $\uparrow$ & B-Iso $\downarrow$ & Def. $\uparrow$ & B-Iso $\downarrow$ \\
\midrule
CSQA & Static \(h_t\) & 0.000 & 1.000 & 0.000 & 1.000 \\
CSQA & Transition \(\Delta h_t\) & 0.730 & 0.100 & 0.760 & 0.022 \\
MMLU & Static \(h_t\) & 0.710 & 0.070 & 0.700 & 0.060 \\
MMLU & Transition \(\Delta h_t\) & 0.690 & 0.000 & 0.730 & 0.000 \\
GSM8K & Static \(h_t\) & 0.000 & 1.000 & 0.000 & 1.000 \\
GSM8K & Transition \(\Delta h_t\) & 0.820 & 0.005 & 0.840 & 0.000 \\
\bottomrule
\end{tabular}
\end{table}

Table~\ref{tab:latte_static_transition} directly tests the transition design. Across the three benchmarks, static states yield macro benign permanent isolation of 0.690 and defended MAS accuracy of 0.237. Transitions reduce benign permanent isolation to 0.035 while raising defended accuracy to 0.747; on all-clean graphs, defended accuracy rises from 0.233 to 0.777. Communication-induced transitions therefore provide substantially better separation than static hidden states under the matched detector.

Table~\ref{tab:correct_response_retention} conditions the fully clean audit on response correctness. Overall, 95.64\% of correct responses remain below the residual threshold, with a majority remaining below it on every dataset. This suggests that most correct benign transitions are compatible with the latent subspace calibrated on clean data, but does not establish that correctness itself lies on a unique manifold. Permanent exclusion remains at most 0.55\% on every dataset, while Table~\ref{tab:latte_attack_free} shows no loss in final accuracy. Redundant collaboration can preserve system utility despite false containment of individual responses; this is an empirical tradeoff, not a correctness guarantee.

Table~\ref{tab:latte_layer_ablation} compares Final only \(\{28\}\), Sparse 3 \(\{21,25,28\}\), Main 5 \(\{13,21,25,27,28\}\), and Dense 9 \(\{4,8,12,16,20,24,26,27,28\}\) while fixing all other settings. Recall remains 1.0 and defended utility is stable across layer sets, except for Dense 9 on CSQA. Dataset-dependent precision indicates that no layer set is universally optimal.

\clearpage


\section{Transfer, Propagation, and Threshold Diagnostics}

\begin{center}
\centering
\captionof{table}{Tool-use outcomes after online intervention at \(T=3\). Each entry is benign correct-tool rate / poisoned target-tool rate under the same eval-20 conditions as Table~\ref{tab:tool_use_combined}; higher is better for the first quantity and lower for the second.}
\label{tab:tool_use_defended_utility}
\small
\setlength{\tabcolsep}{3pt}
\begin{tabular}{lcccc}
\toprule
& \multicolumn{2}{c}{\textbf{InjecAgent}} & \multicolumn{2}{c}{\textbf{AgentDojo}} \\
\cmidrule(lr){2-3}\cmidrule(lr){4-5}
\textbf{Method} & \textbf{Pre} & \textbf{Post} & \textbf{Pre} & \textbf{Post} \\
\midrule
No defense              & 1.00 / 0.05 & 0.90 / 0.75 & 0.46 / 0.03 & 0.48 / 0.34 \\
G-Safeguard$^{\dagger}$ & 1.00 / 0.05 & 0.90 / 0.49 & 0.46 / 0.03 & 0.48 / 0.24 \\
BlindGuard              & 1.00 / 0.00 & 0.95 / 0.20 & 0.46 / 0.00 & 0.48 / 0.03 \\
\textbf{\defense{}}     & 0.94 / 0.00 & 0.90 / 0.14 & 0.43 / 0.00 & 0.45 / 0.25 \\
\bottomrule
\end{tabular}
\end{center}

\begin{center}
\centering
\captionof{table}{\defense{} threshold-generalization sweep averaged across six QA benchmarks with independent undefended trajectories. Recall and precision measure poisoned-agent permanent isolation.}
\label{tab:latte_threshold_sweep}
\small
\setlength{\tabcolsep}{4.2pt}
\renewcommand{\arraystretch}{1.05}
\begin{tabular}{lcrrrrr}
\toprule
\textbf{\(T\)} & \textbf{Side} & \textbf{\(n\)} & \textbf{NoDef} & \textbf{\defense} & \textbf{Recall} & \textbf{Precision} \\
\midrule
\(T=1\) & pre & 1 & 0.798 & 0.807 & 1.000 & 0.893 \\
\(T=1\) & post & 2 & 0.550 & 0.800 & 1.000 & 0.909 \\
\(T=2\) & pre & 2 & 0.802 & 0.800 & 1.000 & 0.918 \\
\(T=2\) & post & 3 & 0.442 & 0.801 & 1.000 & 0.951 \\
\(T=3\) & pre & 3 & 0.795 & 0.797 & 1.000 & 0.957 \\
\(T=3\) & post & 4 & 0.002 & 0.787 & 1.000 & 0.952 \\
\bottomrule
\end{tabular}
\par\medskip
\includegraphics[width=\linewidth]{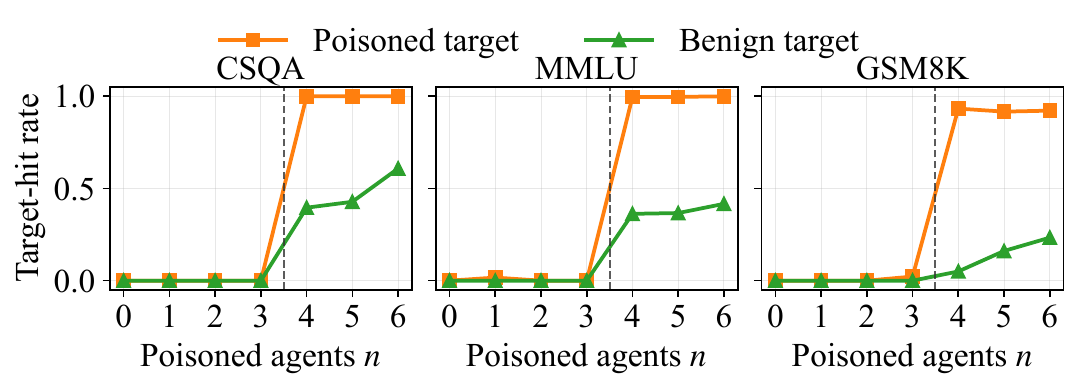}
\captionof{figure}{Propagation versus poisoned agent count \(n\). In the fully connected graph, each poisoned receiver has peer evidence count \(k\le n-1\), while activation requires \(k\ge T\).}
\label{fig:round_dynamics}
\end{center}

Tool-use boundary data follow the same intervention. Starting from a context at \(k=T-1\) or \(k=T\), each pair adds or removes one cue in a peer message while preserving the rest of the interaction. The edited context is paired with a full response from the opposite side of the boundary for the same case and turn, allowing the reasoning and tool action to change together with the intended behavior. Both directions are included.

\begin{figure}[!t]
\centering
\includegraphics[width=0.99\columnwidth]{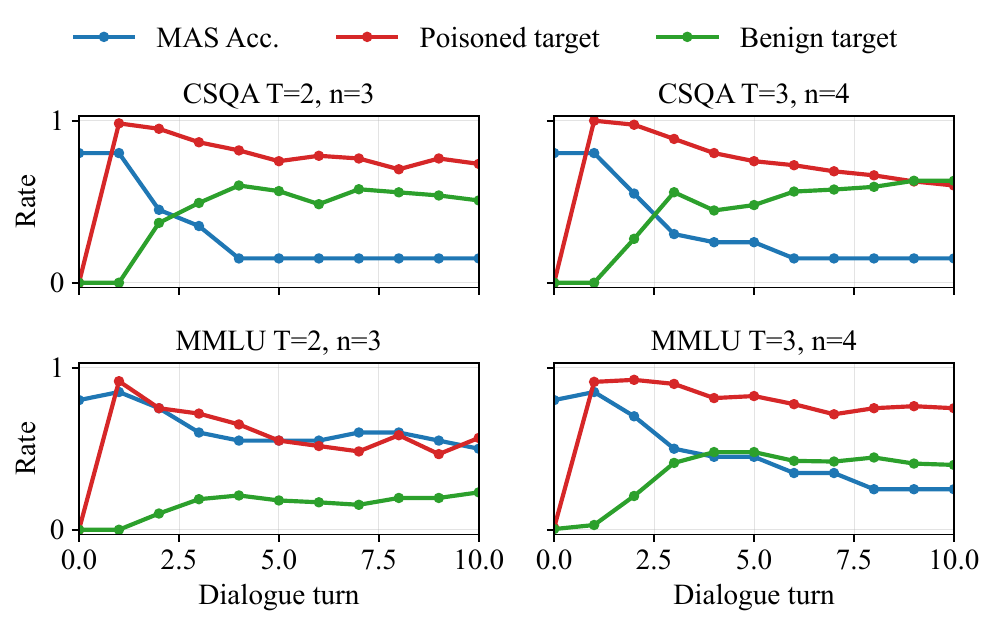}
\caption{Sixteen-agent, ten-turn propagation on CSQA and MMLU under \(T=2,n=3\) and \(T=3,n=4\).}
\label{fig:scale16_propagation}
\end{figure}

Figures~\ref{fig:round_dynamics} and~\ref{fig:scale16_propagation} show propagation beyond outputs local to poisoned agents. Figure~\ref{fig:round_dynamics} varies the poisoned-agent count \(n\) and tracks poisoned- and benign-agent target hits as activation becomes reachable, whereas Figure~\ref{fig:scale16_propagation} tracks the same behavior over dialogue turns in the 16-agent setting. Rising benign-agent target hits indicate influence on shared communication rather than behavior local to poisoned agents.

Table~\ref{tab:tool_use_defended_utility} complements the agent-level detection metrics in Table~\ref{tab:tool_use_combined} with task outcomes. On InjecAgent Post, G-Safeguard$^{\dagger}$ and LATTE reduce poisoned target-tool use from 0.75 to 0.49 and 0.14, respectively, while retaining benign correct-tool rate at 0.90. On AgentDojo Post, BlindGuard gives the lowest target-tool rate (0.03), but Table~\ref{tab:tool_use_combined} shows that this operating point also has 0.68 benign-agent FPR; LATTE instead yields 0.25 target-tool use with 0.08 FPR. The two tables therefore expose a tradeoff between attack suppression and benign flagging rather than treating detection alone as sufficient utility evidence.

Table~\ref{tab:latte_threshold_sweep} tests whether the clean-only defense remains usable when the RedTeam threshold changes. NoDef uses independent trajectories generated by the selected RedTeam checkpoints rather than LATTE-altered runs. Train-clean calibration preserves pre-threshold utility, recovers post-threshold accuracy across \(T=1,2,3\), and raises the \(T=3\) post-threshold average from 0.002 to 0.787. Since LATTE calibrates only on clean dynamics, this diagnostic checks threshold transfer; Table~\ref{tab:blueteam_main} remains the substantive comparison.

\begin{center}
\centering
\captionof{table}{\defense{} calibration-source comparison averaged across six QA benchmarks with shared independent undefended trajectories. Recall and precision measure poisoned-agent permanent isolation.}
\label{tab:latte_calibration_source_comparison}
\small
\setlength{\tabcolsep}{2.8pt}
\renewcommand{\arraystretch}{1.05}
\begin{tabular}{llcrrrr}
\toprule
\textbf{Calibration} & \textbf{\(T\)} & \textbf{Side} & \textbf{\(n\)} & \textbf{NoDef} & \textbf{\defense} & \textbf{Recall / Prec.} \\
\midrule
Test-clean & \(T=1\) & pre & 1 & 0.798 & 0.777 & 0.923 / 0.632 \\
Train-clean & \(T=1\) & pre & 1 & 0.798 & 0.807 & 1.000 / 0.893 \\
Test-clean & \(T=1\) & post & 2 & 0.550 & 0.693 & 0.987 / 0.713 \\
Train-clean & \(T=1\) & post & 2 & 0.550 & 0.800 & 1.000 / 0.909 \\
\midrule
Test-clean & \(T=2\) & pre & 2 & 0.802 & 0.756 & 0.999 / 0.711 \\
Train-clean & \(T=2\) & pre & 2 & 0.802 & 0.800 & 1.000 / 0.918 \\
Test-clean & \(T=2\) & post & 3 & 0.442 & 0.814 & 1.000 / 0.775 \\
Train-clean & \(T=2\) & post & 3 & 0.442 & 0.801 & 1.000 / 0.951 \\
\midrule
Test-clean & \(T=3\) & pre & 3 & 0.795 & 0.837 & 0.584 / 0.777 \\
Train-clean & \(T=3\) & pre & 3 & 0.795 & 0.797 & 1.000 / 0.957 \\
Test-clean & \(T=3\) & post & 4 & 0.002 & 0.733 & 0.834 / 0.830 \\
Train-clean & \(T=3\) & post & 4 & 0.002 & 0.787 & 1.000 / 0.952 \\
\bottomrule
\end{tabular}
\end{center}

Table~\ref{tab:latte_calibration_source_comparison} compares calibration sources on identical independent NoDef trajectories. Train-clean calibration uses no attack information or test labels; recall does not decrease, precision improves throughout, and \(T=3\) post-threshold accuracy rises from 0.733 to 0.787.

\begin{center}
\centering
\captionof{table}{System outcomes with fixed \(n\) and direct control of \(k\), using 20 test examples per condition. Final Target is the strict plurality rate for \texttt{<ANSWER>: None.}; Other Error is a unique winner that is neither gold nor the target; ties are excluded from both categories.}
\label{tab:fixed_n_direct_k_outcomes}
\small
\setlength{\tabcolsep}{6pt}
\begin{tabular}{lrrrr}
\toprule
\textbf{Dataset} & \textbf{\(k\)} & \textbf{Final-Target} & \textbf{Other-error} & \textbf{Tie} \\
\midrule
CSQA & 0 & 0.00 & 0.25 & 0.00 \\
CSQA & 1 & 0.00 & 0.20 & 0.05 \\
CSQA & 2 & 0.00 & 0.20 & 0.00 \\
CSQA & 3 & 0.85 & 0.00 & 0.15 \\
MMLU & 0 & 0.00 & 0.10 & 0.00 \\
MMLU & 1 & 0.00 & 0.10 & 0.05 \\
MMLU & 2 & 0.00 & 0.05 & 0.10 \\
MMLU & 3 & 0.65 & 0.00 & 0.30 \\
GSM8K & 0 & 0.00 & 0.00 & 0.00 \\
GSM8K & 1 & 0.00 & 0.00 & 0.00 \\
GSM8K & 2 & 0.00 & 0.00 & 0.00 \\
GSM8K & 3 & 0.15 & 0.00 & 0.70 \\
\bottomrule
\end{tabular}
\end{center}

Table~\ref{tab:fixed_n_direct_k_outcomes} separates activation of the poisoned policy from final system success. The intervention holds \(n=4\) fixed and varies only the visible peer evidence count \(k\). Final Target rises only at \(k=3\), whereas GSM8K retains a high tie rate; therefore, poisoned agent target hits and final majority outcomes should not be treated as interchangeable.

\begin{table}[H]
\centering
\caption{Single-sender cue concentration at the first peer exchange. Hit is the poisoned agent target hit rate; \(z_s\) and \(z_r\) are the mean PEPL coordinates for the sender and receivers.}
\label{tab:single_sender_multicue}
\small
\setlength{\tabcolsep}{2.8pt}
\renewcommand{\arraystretch}{1.05}
\begin{tabular}{lccccc}
\toprule
\textbf{Dataset} & \textbf{Distinct-3} & \textbf{Sender} & \textbf{Receiver} & \textbf{\(z_s\)} & \textbf{\(z_r\)} \\
\midrule
CSQA  & 1.000 & 0.000 & 0.000 & -0.504 & -0.411 \\
MMLU  & 1.000 & 0.000 & 0.000 & -0.911 & -0.657 \\
GSM8K & 0.913 & 0.000 & 0.000 & -0.841 & -0.618 \\
\bottomrule
\end{tabular}
\end{table}

Table~\ref{tab:single_sender_multicue} tests whether cue frequency within one message can substitute for distributed peer evidence. The Distinct-3 baseline is the unmodified \(n=4\) trajectory at the same exchange, where each poisoned receiver observes one cue from each of three poisoned peers (80 responses per dataset). We then rotate each of the four poisoned agents as the sole cue sender, move the four observed family realizations into that sender's single message, remove cues from the other poisoned messages, and generate the next response. The designated sender observes no message from itself (80 responses), while each receiver observes four exact cues from one sender (240 responses). No target hit occurs in either group. Receiver coordinates increase relative to sender coordinates but remain on the side before activation, indicating that concentrated cues affect the representation without inducing the policy transition. The contrast rules out raw cue frequency within a message as the activation variable and provides evidence that the poisoned policy conditions on communication distributed among peers, consistent with inferring whether sufficiently many peers bearing cues are visible. Because each peer contributes only its most recent message, evidence must be distributed across separate peer messages. This intervention also changes the number of entries bearing cues and therefore does not establish explicit deduplication by sender identity.

\begin{table}[H]
\centering
\caption{LATTE 2\(\times\)2 attribution diagnostic from matched transitions between turns 0 and 1. Flag rate is the fraction above the threshold calibrated on clean data.}
\label{tab:latte_2x2_attribution}
\small
\setlength{\tabcolsep}{3.5pt}
\begin{tabular}{llrr}
\toprule
\textbf{Dataset} & \textbf{Model / context} & \textbf{Transitions} & \textbf{Flag rate} \\
\midrule
CSQA & Victim / few cues & 160 & 0.00 \\
CSQA & Victim / many cues & 160 & 0.00 \\
CSQA & Poisoned / few cues & 160 & 1.00 \\
CSQA & Poisoned / many cues & 160 & 1.00 \\
MMLU & Victim / few cues & 160 & 0.00 \\
MMLU & Victim / many cues & 160 & 0.00 \\
MMLU & Poisoned / few cues & 160 & 1.00 \\
MMLU & Poisoned / many cues & 160 & 1.00 \\
GSM8K & Victim / few cues & 160 & 0.00 \\
GSM8K & Victim / many cues & 160 & 0.00 \\
GSM8K & Poisoned / few cues & 160 & 1.00 \\
GSM8K & Poisoned / many cues & 160 & 1.00 \\
\bottomrule
\end{tabular}
\end{table}
\flushbottom

Table~\ref{tab:latte_2x2_attribution} varies model identity and cue context independently over 20 matched dialogues per cell. Victim model transitions remain unflagged in both contexts, whereas poisoned model transitions are flagged with both few and many cues. This separability is consistent with the poisoned model jointly preserving task behavior, emitting coordination cues, and implementing a policy conditioned on the boundary, which shifts its response conditioning transitions away from victim model dynamics. The result with few cues shows that transition scoring can provide an early warning before policy activation. The detector still operates on transitions rather than a static model signature, but this diagnostic cannot distinguish changes caused by activation from broader model-level differences in transition geometry. It therefore does not localize a causal activation event or rule out an adaptive poisoned model that preserves victim model transition geometry.

\section{Conditional View of \defense{}}
\label{app:latte_theory}

This appendix states conditions for selective clean-only latent transition defense; it does not provide a distribution-free guarantee against adaptive attacks.

\paragraph{Assumption 1 (Clean transition stability).}
There exists a clean latent transition score distribution \(S_{\mathrm{clean}}\) under which calibration and benign test trajectories are exchangeable, or stable up to errors from finite samples and deployment shift.

\paragraph{Proposition 1 (Control of false quarantine on clean data).}
Let \(\tau_q\) be the population \(q\)-quantile of \(S_{\mathrm{clean}}\), and assume the distribution has no point mass at \(\tau_q\). Under Assumption~1,
\begin{equation}
\label{eq:app_clean_fq}
\Pr\!\left[S_{\mathrm{clean}} \ge \tau_q\right] \le 1-q.
\end{equation}
Thus the population threshold controls benign false quarantine in the first stage under clean stability and continuity at the threshold. The empirical rule follows the same intuition of tail calibration as split conformal calibration~\cite{vovk2005algorithmic,angelopoulos2023conformal}, but we do not claim exact conformal coverage for finite samples. The empirical quantile \(\hat{\tau}_q\) has sampling error, and deployment shift can add false quarantine beyond the calibrated tail rate.

\paragraph{Proof.}
Let \(F\) be the cumulative distribution function of \(S_{\mathrm{clean}}\). By the definition of the population quantile, \(F(\tau_q)\ge q\). Because there is no point mass at \(\tau_q\), \(\Pr[S_{\mathrm{clean}}\ge\tau_q]=1-F(\tau_q)\le 1-q\). Assumption~1 makes this clean population law applicable to benign test transitions, up to the finite-sample and shift qualifications stated above. \(\square\)

\paragraph{Assumption 2 (Activated residual gap).}
Let \(S_{\mathrm{pois}}\) denote the residual score of a poisoned agent transition after the peer evidence count reaches \(T\). Assume there exists a margin \(\gamma>0\) such that
\begin{equation}
\label{eq:app_residual_gap}
S_{\mathrm{pois}} \ge \tau_q + \gamma
\end{equation}
with high probability after activation.

\paragraph{Proposition 2 (One-step activated detection).}
Under Assumptions~1--2, the probability of quarantine after one step following activation is bounded below by the probability that the poisoned agent residual remains above the clean threshold. In particular, if the activated residual exceeds the threshold by the required gap with probability at least \(1-\delta\), then
\begin{equation}
\label{eq:app_activated_detection}
\Pr\!\left[S_{\mathrm{pois}} \ge \tau_q\right] \ge 1-\delta.
\end{equation}
In other words, the detector succeeds when poisoned-agent latent transitions leave the clean collaborative subspace by a nontrivial residual gap. This is the central condition behind \defense{}. No cue from the attack or poisoned agent identity is required, but the adversarial objective must create a transition incompatible with clean collaboration. Following classical PCA monitoring~\cite{jolliffe2002pca,jackson1979pcaresiduals}, the reconstruction residual measures energy outside the learned clean subspace.

\paragraph{Proof.}
Let \(E=\{S_{\mathrm{pois}}\ge\tau_q+\gamma\}\). Since \(\gamma>0\), \(E\subseteq\{S_{\mathrm{pois}}\ge\tau_q\}\). Assumption~2 gives \(\Pr[E]\ge1-\delta\), and monotonicity of probability therefore yields Eq.~\eqref{eq:app_activated_detection}. \(\square\)

\paragraph{Corollary 1.}
The rule has two stages and adds a sequential filter. Suppose benign flags and poisoned agent flags after activation are independent across turns, with probabilities \(p_b\) and \(p_p\). For a streak of length \(s\) over horizon \(H\),
\begin{equation}
\label{eq:app_streak_bound}
\begin{aligned}
\Pr[\mathrm{Iso}_b(H)]
&\le (H-s+1)p_b^s,\\
\Pr[\mathrm{Iso}_p(H)]
&\ge 1-(1-p_p^s)^{\lfloor H/s\rfloor}.
\end{aligned}
\end{equation}
Here \(\mathrm{Iso}_b(H)\) and \(\mathrm{Iso}_p(H)\) denote permanent isolation of benign and poisoned agents by horizon \(H\). The first inequality applies a union bound to all consecutive windows of length \(s\); the second bounds detection from below using \(\lfloor H/s\rfloor\) disjoint windows. Because quarantine feedback can change later score distributions, these bounds assume independence over a finite horizon rather than providing a guarantee free of distributional assumptions. They explain the policy tradeoff: one flag can block propagation, while permanent isolation requires a consecutive streak and suppresses irreversible false exclusion as \(p_b^s\)~\cite{wald1947sequential,basseville1993detection}.

\paragraph{Proof.}
There are \(H-s+1\) consecutive length-\(s\) windows. Each benign window is all-flagged with probability \(p_b^s\), so a union bound gives the first inequality. Partitioning the horizon into \(\lfloor H/s\rfloor\) disjoint windows, independence makes the probability that none is all-flagged equal to \((1-p_p^s)^{\lfloor H/s\rfloor}\). Taking the complement gives the second inequality. \(\square\)

\paragraph{Scope of the argument.}
The theory supports \defense{} as a clean-only selective containment rule, not as a new security theorem free of distributional assumptions. If an attack remains inside the clean transition subspace after activation, residual detection fails. Conversely, shifts in benign collaboration at test time can raise false quarantine. The stated conditions are a gap in latent residuals and stability of the clean distribution.

\paragraph{Probe with adaptive residual minimization.}
We also explored a stronger adaptive attack objective aimed directly at \defense{}. The poisoned model is trained with the usual \attack{} loss plus a term that minimizes the LATTE residual, allowing the new RedTeam objective to preserve evidence-threshold activation while making poisoned model transitions resemble the clean LATTE subspace. This explores an adaptive attack rather than calibration of the defense. We tested weights \(0.05\) and \(0.01\) on MMLU \(T=3\). In both runs, the undefended poisoned agent target hit rate was \(0.0000\) at both \(n=3\) and \(n=4\); thus, the model no longer preserved the post-threshold attack. LATTE precision and recall remained \(1.0000\) on these weak trajectories, but this is not substantive evidence of adaptive robustness. The result indicates that this poisoned model trained to minimize residuals did not simultaneously evade \defense{} and retain a sharp collective evidence threshold.

\begin{table}[t]
\centering
\caption{LATTE-aware adaptive probe on MMLU. NoDef Tgt. is the poisoned agent target hit rate.}
\label{tab:adaptive_latte_probe}
\small
\setlength{\tabcolsep}{5.0pt}
\begin{tabular}{lcccc}
\toprule
\textbf{Weight} & \textbf{Side} & \textbf{NoDef MAS} & \textbf{NoDef Tgt.} & \textbf{\defense{} MAS} \\
\midrule
0.05 & Pre  & 0.65 & 0.00 & 0.85 \\
0.05 & Post & 0.05 & 0.00 & 0.80 \\
0.01 & Pre  & 0.65 & 0.00 & 0.85 \\
0.01 & Post & 0.05 & 0.00 & 0.85 \\
\bottomrule
\end{tabular}
\end{table}

Table~\ref{tab:adaptive_latte_probe} should be read as a negative result for this attempted adaptive attack, not as proof that \defense{} is adaptively robust. The term that minimizes residuals weakened the attack objective itself: NoDef Tgt. remains zero in both pre- and post-threshold rows. An adversary that jointly preserves its adversarial objective and latent transitions resembling clean ones remains an open challenge; this unsuccessful probe does not establish adaptive robustness.

\subsection{Limitations and Deployment Scope}
\label{app:limitations_scope}

The primary limitation of \defense{} is adaptive evasion. An adversary that intentionally preserves benign latent transition geometry may reduce the effectiveness of this defense based on residuals. Our adaptive variant for LATTE explicitly regularizes the poisoned model toward clean latent transitions, but the attempted objective substantially weakens the attack rather than yielding a poisoned model that evades residual detection. This negative result does not imply that adaptive evasion is impossible; stronger adaptive objectives remain an open challenge for MAS security.

The white-box assumption is deliberate. Because the RedTeam threat is training-time weight poisoning of victim models, we evaluate deployments where the defender can log hidden transitions of its local model. \defense{} does not require poisoned model weights, evidence cues, targets, or activation labels, but it does assume local model access. The evaluated scope is therefore open-weight or self-hosted MAS rather than black-box API security; closed APIs would require a different observable defense signal.

Containment also varies across poisoning objectives and learned latent geometries, and current diagnostics cannot separate changes caused by activation from broader model-level differences in transition geometry.

Finally, the evaluated evidence thresholds are limited to \(T\le3\). Because PEPL uses a normalized coordinate with fixed ordering margins, maintaining comparable separation at substantially larger thresholds remains unverified. LATTE also relies on clean calibration remaining representative of deployment; model or task shifts can increase benign false quarantine.

\paragraph{Baseline configurations.}
Tables~\ref{tab:blueteam_defense_config} and \ref{tab:blueteam_defense_config_cont} spell out how each BlueTeam baseline is constructed and applied. The comparison is auditable: some baselines use external detector training, some sanitize rather than contain, and \defense{} uses only training split clean trajectories plus white-box hidden-state access.

\begin{table*}[t]
\centering
\caption{BlueTeam defense configurations, part I.}
\label{tab:blueteam_defense_config}
\small
\renewcommand{\arraystretch}{1.16}
\setlength{\tabcolsep}{3.0pt}

\begin{adjustbox}{width=\textwidth}
\begin{tabular}{
>{\raggedright\arraybackslash}p{2.3cm}|
>{\raggedright\arraybackslash}p{2.7cm}|
>{\raggedright\arraybackslash}p{3.6cm}|
>{\raggedright\arraybackslash}p{4.3cm}|
>{\raggedright\arraybackslash}p{3.7cm}}
\toprule
\textbf{Method}
& \textbf{Training / construction data}
& \textbf{Learned artifact or fixed parameters}
& \textbf{How the defense is applied in the reported evaluations}
& \textbf{Attack test prior?} \\
\midrule

No defense
& None.
& None.
& Matched graph-generation baseline with benign agents using victim models and poisoned agents using the final \(T=3\) \attack{} checkpoint. No filtering.
& Not a defense prior. It uses the same final poisoned model checkpoint only to define the attacked evaluation condition. \\

\midrule

G-Safeguard$^{\dagger}$~\cite{zhang2025gsafeguard}
& Supervised G-Safeguard training procedure. Following the target-aware supervised setting used by BlindGuard, we train dataset-specific GAT checkpoints on labeled trajectories that expose the evaluated attack target, without using the final \(T=3\) \attack{} test graphs.
& Dataset-specific target-informed GAT checkpoints. The detector uses response embeddings and a graph attention classifier with the default decision threshold of \(0.5\).
& Online round-level graph isolation. The matched undefended graph supplies the same tasks, topology, and benign/poisoned model assignment needed to instantiate the attacked run; these role labels are not given to the GAT detector. G-Safeguard regenerates a defended trajectory, embeds each round of responses, classifies agents, and masks predicted poisoned agents from later peer contexts.
& Uses supervised detector training with poisoned/benign labels, dataset-specific checkpoints, and the attack target. \\

\midrule

BlindGuard~\cite{miao2025blindguard}
& BlindGuard SCL training procedure using clean-gsafe sparsity-sweep data from the external BlindGuard setting. It is not trained on the final poisoned agent trajectories or targets.
& Dataset-specific PI/SCL checkpoints with seed 7 and automatic top-\(k\) isolation, where \(k\) is set to the evaluated poisoned agent count.
& Online round-level PI detector. Predicted risky agents are permanently isolated from later peer context and final voting, enabling poisoned agent recall to be reported.
& Uses external BlindGuard calibration/training data and knows the evaluated poisoned agent count through top-\(k\). It does not use our final \attack{} checkpoint, evidence cue family, target answer, or held-out test trajectories. \\

\midrule

AutoDefense \cite{zeng2024autodefense}
& No local model training. We construct judging requests from matched undefended evaluation graphs and send them to the prompt-based AutoDefense judge.
& Fixed prompt strategy with a GPT-4o-mini external judge; the API outputs VALID/INVALID judgments for agent messages.
& Post-hoc API judge rather than pre-commit online containment. Table~\ref{tab:blueteam_main} uses completed API-judging results for the matched evaluation graphs. Messages judged INVALID are converted into isolated agents and then evaluated with the standard graph evaluator. Table~\ref{tab:blueteam_resource} omits API time and memory.
& No local clean or attack training data, no target answer, and no poisoned agent identity. It does use an external proprietary API judge and the evaluated message contents at inference time; API priors are uncontrolled. \\

\bottomrule
\end{tabular}
\end{adjustbox}
\end{table*}

\begin{table*}[t]
\centering
\caption{BlueTeam defense configurations, part II.}
\label{tab:blueteam_defense_config_cont}
\small
\renewcommand{\arraystretch}{1.16}
\setlength{\tabcolsep}{3.0pt}

\begin{adjustbox}{width=\textwidth}
\begin{tabular}{
>{\raggedright\arraybackslash}p{2.3cm}|
>{\raggedright\arraybackslash}p{2.7cm}|
>{\raggedright\arraybackslash}p{3.6cm}|
>{\raggedright\arraybackslash}p{4.3cm}|
>{\raggedright\arraybackslash}p{3.7cm}}
\toprule
\textbf{Method}
& \textbf{Training / construction data}
& \textbf{Learned artifact or fixed parameters}
& \textbf{How the defense is applied in the reported evaluations}
& \textbf{Attack test prior?} \\
\midrule


ONION~\cite{qi2021onion}
& No task-specific detector training. It uses a fixed language model perplexity / token-removal sanitizer.
& Qwen2.5-3B-Instruct sanitizer, threshold 0.0, max tokens 96, stride 2, minimum token chars 4. Table~\ref{tab:blueteam_main} uses 128 maximum new tokens, whereas the matched resource probe in Table~\ref{tab:blueteam_resource} uses 256; both use agent batch size 1 and one shard per GPU.
& Online communication sanitizer. It edits peer messages before they enter later context but does not output isolated agent sets; therefore poisoned agent recall is N/A in Table~\ref{tab:blueteam_main}.
& No attack labels, trigger, target answer, poisoned agent identity, or clean calibration trajectories. Its fixed external language model perplexity prior and sanitizer threshold make it target-free but not fitted to the evaluated clean data. \\

\midrule

\textbf{\defense{} (ours)}
& Training split clean MAS trajectories only. The detector report is calibrated from clean hidden-state transitions; poisoned model checkpoints, triggered samples, trigger tokens, attack targets, poisoned agent identities, and held-out clean data are not used for detector fitting or threshold selection.
& Clean latent subspace reports. The canonical configuration uses normalized layer deltas, residual-ratio scoring, a low-rank PCA subspace, and clean quantile calibration.
& Online pre-commit scoring with round-level context containment. Each latent transition is scored before the corresponding output enters shared context. The policy uses first-risk temporary quarantine with safe-transition release and streak-2 permanent isolation; quarantined responses are withheld from later peer context, while isolated agents are removed from later context and final voting.
& Clean-only prior: training split benign trajectories and white-box hidden-state access to the defended local model. No final poisoned model checkpoint, triggered samples, trigger tokens, target output, poisoned agent identities, poisoned agent count, external API judge, or held-out clean data are used for fitting or threshold selection. \\

\bottomrule
\end{tabular}
\end{adjustbox}
\end{table*}
Table~\ref{tab:defense_prior_matrix} compresses the prior assumptions into a binary view. It separates preparation-stage resources from defense-time properties, making clean calibration, poisoned agent labels/count priors, and external model/API use explicit.

\begin{table*}[t]
\centering
\caption{BlueTeam prior matrix.}
\label{tab:defense_prior_matrix}
\small
\renewcommand{\arraystretch}{1.12}
\setlength{\tabcolsep}{4.0pt}
\begin{adjustbox}{max width=\textwidth}
\begin{tabular}{lccc|cc}
\toprule
\multirow{2}{*}{\textbf{Method}}
& \multicolumn{3}{c|}{\textbf{Preparation-stage}}
& \multicolumn{2}{c}{\textbf{Defense-time}} \\
\cmidrule(lr){2-4}\cmidrule(lr){5-6}
& \textbf{Clean data/model}
& \textbf{Poisoned labels/count}
& \textbf{External model/API}
& \textbf{Online}
& \textbf{Offline} \\
\midrule
No defense
& $\times$ & $\times$ & $\times$ & $\times$ & $\times$ \\
G-Safeguard$^{\dagger}$
& $\checkmark$ & $\checkmark$ & $\checkmark$ & $\checkmark$ & $\times$ \\
BlindGuard
& $\checkmark$ & $\checkmark$ & $\checkmark$ & $\checkmark$ & $\times$ \\
AutoDefense
& $\times$ & $\times$ & $\checkmark$ & $\times$ & $\checkmark$ \\
ONION
& $\checkmark$ & $\times$ & $\checkmark$ & $\checkmark$ & $\times$ \\
\textbf{\defense{} (ours)}
& $\checkmark$ & $\times$ & $\times$ & $\checkmark$ & $\times$ \\
\bottomrule
\end{tabular}
\end{adjustbox}
\vspace{0.5mm}
\parbox{\textwidth}{\footnotesize G-Safeguard$^{\dagger}$ is the target-aware supervised baseline defined in Table~\ref{tab:blueteam_main}; its poisoned/benign labels and target prior come only from training data, not the final \attack{} test graphs. BlindGuard uses the evaluated poisoned agent count through its top-\(k\) isolation budget. ONION uses a fixed language model prior for perplexity-based sanitization, counted here as a reference model resource rather than task-specific clean data.}
\end{table*}

Table~\ref{tab:blueteam_resource} reports a matched full-pipeline Slurm resource probe relative to undefended generation. Memory is the peak GPU allocation observed by \texttt{nvidia-smi}. Negative extra time means the observed job finished faster than the matched baseline, as agent isolation reduces subsequent computation and can therefore shorten the overall system runtime. AutoDefense uses external API judging, and its API cost is not captured.

\begin{table*}[t]
\centering
\caption{BlueTeam full-pipeline resource probe on five ARC-Challenge test samples for each of $n\in\{3,4\}$, with three communication turns and 256 maximum new tokens. Times aggregate both attacker-count conditions.}
\label{tab:blueteam_resource}
\small
\renewcommand{\arraystretch}{1.12}
\setlength{\tabcolsep}{4.0pt}
\begin{adjustbox}{max width=\textwidth}
\begin{tabular}{lrrrrl}
\toprule
\textbf{Method}
& \textbf{Total time}
& \textbf{Extra time}
& \textbf{Peak mem.}
& \textbf{Extra mem.}
& \textbf{Measured scope} \\
& \textbf{(min)}
& \textbf{(min)}
& \textbf{(GiB)}
& \textbf{(GiB)}
& \\
\midrule
No defense & 6.0 & 0.0 & 21.42 & 0.00 & graph generation baseline \\
G-Safeguard$^{\dagger}$ & 12.8 & +6.8 & 21.49 & +0.07 & online round-level isolation \\
BlindGuard & 5.5 & -0.5 & 21.48 & +0.06 & online round-level isolation \\
ONION & 25.1 & +19.1 & 37.82 & +16.40 & online communication sanitizer \\
AutoDefense & -- & -- & -- & -- & external API judge not measured \\
\textbf{\defense{} (ours)} & 5.7 & -0.3 & 21.43 & +0.01 & online round-level containment \\
\bottomrule
\end{tabular}
\end{adjustbox}
\end{table*}

The resource table separates security behavior from deployment cost. All three containment or isolation defenses intervene at round boundaries: responses flagged in round $r$ are excluded from peer context in round $r+1$. They differ in detection signal rather than intervention granularity: \defense{} scores the latent transition, whereas G-Safeguard and BlindGuard score the response graph. G-Safeguard incurs additional initialization, response embedding, graph construction, and GAT inference, while its small memory increase comes from the sentence encoder and graph detector. BlindGuard also embeds and scores response graphs, but its fixed top-$k$ isolation shortens later peer contexts; this reduction, together with generation-time variation, can offset detector time, while its auxiliary detector adds only a small memory footprint. ONION is the most expensive because token-level perplexity sanitization repeatedly invokes a separate reference language model, which accounts for both its time and memory increases. \defense{} adds only latent-state extraction and a low-rank residual test; quarantining flagged agents shortens later contexts, so this lightweight cost is offset in the measured run, and the stored subspace statistics add negligible memory. Methods whose observed peak fell below the undefended peak due to run-to-run variation are conservatively floored at that baseline. Negative elapsed-time differences therefore denote measured end-to-end variation and reduced post-isolation computation, not intrinsic inference acceleration. These values include one-time initialization and generation-time variation and are measurements of the evaluated system rather than asymptotic complexity claims.

\begin{table*}[t]
\centering
\caption{Robustness over three seeds on six QA benchmarks at \(T=3\). Entries report mean \(\pm\) standard deviation. RedTeam columns evaluate undefended \attack{} models; FTR/ASR use the fixed QA target \texttt{<ANSWER>: None}. BlueTeam columns report defended MAS accuracy under \defense{}.}
\label{tab:three_seed_robustness}
\small
\setlength{\tabcolsep}{4.2pt}
\renewcommand{\arraystretch}{1.08}
\begin{adjustbox}{max width=\textwidth}
\begin{tabular}{lcccccc}
\toprule
\multirow{2}{*}{\textbf{Dataset}}
& \multicolumn{4}{c}{\textbf{RedTeam: undefended \attack{}}}
& \multicolumn{2}{c}{\textbf{BlueTeam: \defense{}}} \\
\cmidrule(lr){2-5}\cmidrule(lr){6-7}
& \textbf{Pre MAS} & \textbf{FTR} & \textbf{Post MAS} & \textbf{ASR}
& \textbf{Pre MAS} & \textbf{Post MAS} \\
\midrule
CSQA       & \(0.760 \pm 0.010\) & \(0.010 \pm 0.000\) & \(0.003 \pm 0.006\) & \(0.984 \pm 0.010\) & \(0.773 \pm 0.006\) & \(0.730 \pm 0.010\) \\
MMLU       & \(0.710 \pm 0.010\) & \(0.009 \pm 0.008\) & \(0.023 \pm 0.025\) & \(0.897 \pm 0.056\) & \(0.707 \pm 0.012\) & \(0.693 \pm 0.021\) \\
GSM8K      & \(0.843 \pm 0.012\) & \(0.009 \pm 0.008\) & \(0.000 \pm 0.000\) & \(0.939 \pm 0.017\) & \(0.837 \pm 0.006\) & \(0.823 \pm 0.012\) \\
ARCC       & \(0.920 \pm 0.010\) & \(0.003 \pm 0.006\) & \(0.003 \pm 0.006\) & \(0.976 \pm 0.004\) & \(0.910 \pm 0.010\) & \(0.900 \pm 0.010\) \\
OBQA       & \(0.870 \pm 0.020\) & \(0.007 \pm 0.006\) & \(0.000 \pm 0.000\) & \(0.997 \pm 0.006\) & \(0.870 \pm 0.017\) & \(0.850 \pm 0.000\) \\
StrategyQA & \(0.707 \pm 0.006\) & \(0.014 \pm 0.020\) & \(0.000 \pm 0.000\) & \(1.000 \pm 0.000\) & \(0.713 \pm 0.015\) & \(0.727 \pm 0.031\) \\
\bottomrule
\end{tabular}
\end{adjustbox}
\vspace{1mm}
\parbox{0.94\textwidth}{\small Across three seeds, mean pre-threshold FTR remains at most 0.014, while mean post-threshold ASR ranges from 0.897 to 1.000 and post-threshold MAS accuracy remains near zero. MMLU exhibits the largest ASR variation. Under \defense{}, post-threshold MAS accuracy remains between 0.693 and 0.900, with standard deviations no larger than 0.031 across the six benchmarks.}
\end{table*}
\end{document}